\documentclass[pre,aps,twocolumn,showpacs,floatfix,superscriptaddress]{revtex4-1}
\usepackage{graphicx}% Include figure files only as boxes for faster compilation- add [draft] option here
\usepackage{dcolumn}% Align table columns on decimal point
\usepackage{bm}% bold math
\usepackage{hyperref}% add hypertext capabilities
\usepackage{color}
\usepackage{epsf,epsfig}
\usepackage{color}
\usepackage{amsmath}
\usepackage{enumerate}
\usepackage{tabularx}
\usepackage{array}
\usepackage{xcolor}
\usepackage[normalem]{ulem}

\usepackage{amssymb}
\usepackage{subfigure}
\usepackage{lineno}

\begin{document}
\title{Learning Associations in Reconfigurable Particle Packings via Local Cyclic Driving}
\author{Wenjing Guo}
\thanks{These two authors contributed equally.}
\affiliation{School of Chemistry and Chemical Engineering, Shanghai Jiao Tong University, Shanghai 200240, China}
\author{Vidyesh Rao Anisetti}
\thanks{These two authors contributed equally.}
\affiliation{James Franck Institute, University of Chicago, Chicago IL, USA}
\author{Kairui Zhang}
\affiliation{School of Chemistry and Chemical Engineering, Shanghai Jiao Tong University, Shanghai 200240, China}
\author{Shabeeb Ameen}
\affiliation{Physics Department, Syracuse University, Syracuse, NY 13244 USA}
\author{Ananth Kandala}
\affiliation{Physics Department, University of Florida, Gainesville, FL 32611 USA }
\author{Menachem Stern}
\affiliation{AMOLF, Science Park 104, 1098 XG Amsterdam, The Netherlands}
\author{Nidhi Pashine}
\affiliation{Physics Department, Syracuse University, Syracuse, NY 13244 USA}
\author{Joseph D. Paulsen}
\affiliation{Physics Department, Syracuse University, Syracuse, NY 13244 USA}
\affiliation{Department of Physics, St.~Olaf College, Northfield, MN 55057 USA}
\author{J. M. Schwarz}
\email{jmschw02@syr.edu}
\affiliation{Physics Department, Syracuse University, Syracuse, NY 13244 USA}
\affiliation{Indian Creek Farm, Ithaca, NY 14850 USA}
\author{Tao Zhang}
\email{zhangtao.scholar@sjtu.edu.cn}
\affiliation{School of Chemistry and Chemical Engineering, Shanghai Jiao Tong University, Shanghai 200240, China}

\date{\today}

\begin{abstract}
We investigate \emph{associative-memory} behavior in a reconfigurable particle packing programmed by purely local cyclic driving. The system is a two-dimensional bidisperse Lennard--Jones particle assembly with periodic boundaries evolved under athermal quasistatic relaxation. During training, a fixed set of input particles is driven cyclically while output particles are selected on-the-fly by a region-driving rule and driven according to a prescribed flow  pattern; during retrieval, only the inputs are driven. Associative-memory performance is quantified by the cosine similarity between the realized and target output displacement directions.
Unlike physical learning systems with fixed architecture, learning here arises through emergent weight updates: localized rearrangements modify the contact network and thereby reshape the effective mechanical couplings between the inputs and the outputs. Across task difficulty we identify three distinct regimes. In an \emph{easy} setting, the material's intrinsic mechanical response already produces coherent, same-direction motion of the right-hand region under input-only driving, yielding high associative-memory performance without training; driving outputs during training perturbs this favorable response and slightly degrades performance. In a \emph{hard} setting, the target mapping requires output motions that are incompatible with the dominant collective drift, resulting in low baseline performance and only modest training gains; introducing intermittent relaxation cycles during training reduces the train-- mismatch and measurably improves associative-memory performance. In an intermediate \emph{quadrupolar} task, repositioning the input--output geometry robustly locks in the desired response and converts initially stochastic trajectories into reproducible learned motions.
Finally, under quadrupolar geometric driving we perform parameter sweeps that reveal practical control knobs for programmability---including staged ramping of drive amplitude, gradual tuning of the interaction length scale, and controlled particle-size heterogeneity with input amplification---that regulate learnability and the emergence of stable periodic orbits. Together, these results identify minimal physical ingredients for association-based functionality in athermally driven particulate media and motivate an association learning phase diagram for reconfigurable matter.

%association-based functionality versus write-store-retrieve functionality
\end{abstract}

\maketitle

\section{Introduction}
\label{sec_introduction}

Many physical systems are capable of transmitting information from one location to another through their internal structure. In mechanical networks, this phenomenon can give rise to allostery: a localized perturbation applied at one site induces a specific, reproducible response elsewhere in the material. Inspired by biological allostery with proteins~\cite{Tirion1996,Atilgan2001,Gordon2015Notch}, recent work has shown that spring networks and mechanical metamaterials can be trained to exhibit such targeted input–output relationships by tuning internal couplings such as spring constants or rest lengths~\cite{rocks2017designing,Yan2017PNAS,Coulais2018MultiStep}. In these allosteric systems, the training task is fundamentally relational—a particular pattern of inputs should reliably elicit a particular pattern of outputs. In addition to allostery, researchers have been able to train elastic networks to achieve specific bulk properties, such as the Poisson's ratio or the nonlinear elastic response, in a process called directed aging \cite{pashine2019directed,hexner2020effect}. These feats demonstrate the trainability of mechanical networks.

Beyond training for allostery and bulk material properties, there has been a growing interest in the broader notion of physical learning, where physical networks can learn by updating parameters, similar to updating weights in artificial neural networks~\cite{scellier2017equilibrium,stern2021supervised,anisetti2023,anisetti2024frequency,falk2025temporal,stern2023learning,momeni2025training}. These physical networks can learn in the generalization sense of the word in terms of the learning being applied to new inputs. This notion is not just  theoretical. Experimental examples to date include resistor networks and transistor networks, which can be trained by updating resistances~\cite{dillavou2022demonstration,Dillavou2024PNAS}, and elastic networks, which can be trained by updating rest lengths and stiffness through training protocols~\cite{Altman2024,lee2022mechanical,li2024training}. In such systems, the updates are given by a particular rule and typically performed by a supervisor, be it a person or a computer. And while this paradigm shows great promise, much, though not all, of the effort presumes that the topology of interactions is static, which is a strong assumption that is simply not the case in many materials.

Indeed many disordered systems of interest, including particle packings, foams, and amorphous solids, exemplify this dual nature~\cite{FalkLanger1998,MaloneyLemaitre2006,KablaDebregeas2003,UtterBehringer2004}. At sufficiently high packing fractions, such materials can transmit stresses over long distances through heterogeneous force networks, yet they remain highly susceptible to localized rearrangements under external driving. Under cyclic or quasistatic loading, their response often proceeds through discrete, history-dependent rearrangements, endowing the system with memory of past deformations~\cite{Keim2019memory}. As a result, the mechanical response of such systems is not determined solely by the instantaneous load, but by the entire history of applied driving \cite{Paulsen25mechanical}. This history dependence suggests a natural abstract description: the set of mechanically stable configurations of the material, together with the transitions between them induced by driving, form a state-transition structure reminiscent of a finite state machine \cite{Mungan2019networks,Mungan2019cyclic,Paulsen2019minimal,keim2021multiperiodic,Liu2024}. From this perspective, a driving protocol acts as an input, while the resulting sequence of rearrangements constitutes the output. Although we do not explicitly identify or manipulate this state-transition graph here, this viewpoint highlights an important conceptual distinction between reconfigurable materials and fixed-architecture networks. In reconfigurable systems, memory is not necessarily stored solely in continuously tunable parameters, but can also arise from structural rearrangements that reshape the material’s internal organization. In the delicate balance between reconfigurability and structural rigidity; the former allows for adaptation, while the latter ensures that encoded memories remain stable.

Here, we focus on a fundamental and experimentally relevant question within this broader landscape: can a reconfigurable particle packing be trained to form stable associations between specific inputs and specific output responses much like neurons in the brain form associations between distinct events? Using a two-dimensional packing subjected to purely local cyclic driving, we show that such systems can indeed be trained to exhibit associative-memory-like behavior in which specific input motions reliably create a partial pattern that can be used to retrieve targeted output motions just as partial Hopfield model firing patterns were used to retrieve the rest of the firing pattern~\cite{hopfield1982neural}. By systematically varying task geometry and training protocols, we reveal that not all associative tasks are equally learnable. Specifically, we find that associative learning in reconfigurable matter naturally organizes into easy, hard, and intermediate regimes set by the compatibility between the target input–output relation and the system’s intrinsic mechanical response. This classification highlights that learning in reconfigurable matter is not solely a matter of training protocol, but is fundamentally constrained by the system’s internal mechanics.

While our results focus on associative memory rather than explicit state-machine learning, they point toward a broader implication: in reconfigurable matter, learning need not be restricted to modifying effective weights on a fixed graph, but can instead arise from structural plasticity that reshapes the material’s internal organization. In this sense, reconfiguration itself can play a role analogous to adaptive weights in neural networks, providing a physical mechanism for updating effective interactions without explicit parameter tuning. In other words, the tuning of the weights is itself an emergent, unsupervised process.  We return to the implications of this perspective in the Discussion, where we outline how associative learning in particle packings may serve as a stepping stone toward training the state-transition structures themselves.

\section{Training Algorithm}
\label{sec_methodology}
\begin{figure*}[h]
	\centering
	\includegraphics[width=0.9\textwidth]{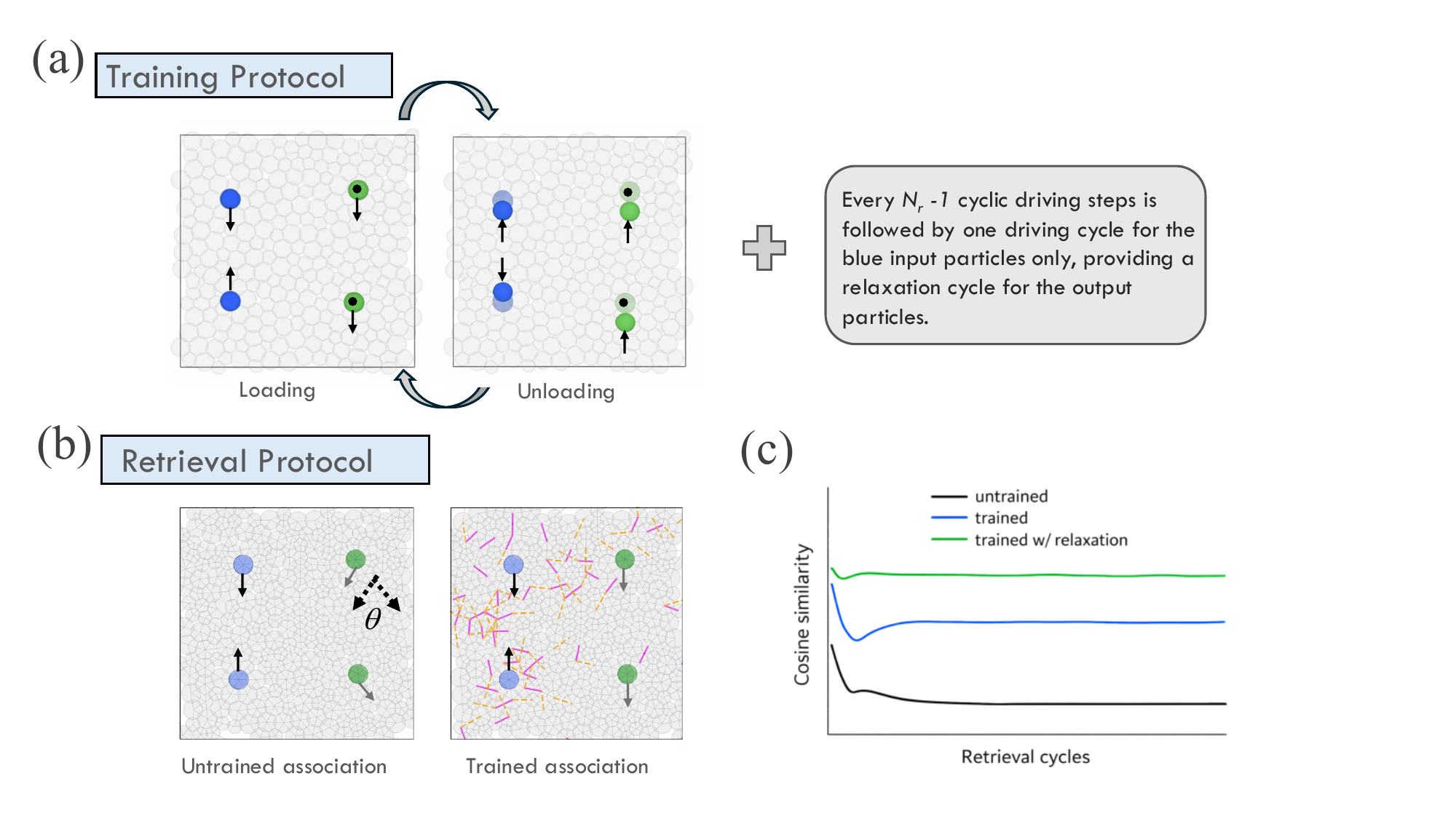}
	\caption{{\it Schematic of the training protocol, the retrieval protocol, and a performance metric.} (a) The task is to associate two input particles  (blue) with two output particles (green) to create a particular flow pattern in terms of direction. For this specific task, when the two blue input particles move closer together vertically, the green output particles both move downward. The training protocol is such that there are $N_r-1$ loading/unloading cycles in which both the inputs and the outputs are cyclically driven. In the unloading schematic, the lighter shading represent the initial positions of the particles. After $N_r-1$ loading/unloading cycles, for some tasks, a relaxation cycle is implemented in which the output particles are not driven. This process constitutes one training cycle. In addition, the filled black circles denotes that we choose two output coordinates initially and particles closest to the two coordinates are chosen as the output particles after each cycle. In other words, the association is not specifically between two chosen input particles and two chosen output particles but rather the particles nearest to the two output regions given the potential rearrangements and that we are training for direction of displacement and not magnitude (or displacement). 
(b) To retrieve the association in the form of a flow pattern, the black arrows denoting the direction of driving for the loading stage of the cycle and the grey arrows denote the direction of displacement (without driving). Simply put, only the input particles are driven in the retrieval stage. The magenta edges represent contacts gained during a training cycle and the dashed orange edges lost during a training cycle. Other contact edges are shown in lighter grey. 
(c) Schematic plot of the cosine similarity, or the cosine of the angle denoted in (b) as a function of the retrieval cycle, which is our performance measure.}

\label{fig:protocol}
\end{figure*}
\subsection{Model Implementation and Particle System}
We simulate a two-dimensional assembly of $N$ bidisperse soft spheres confined within a square box of side length $L$ with periodic boundary conditions. The particles interact via a Lennard-Jones potential,
\begin{equation}
V(r_{ij}) = 4\epsilon \left[ \left(\frac{\sigma_{ij}}{r_{ij}}\right)^{12} - \left(\frac{\sigma_{ij}}{r_{ij}}\right)^6 \right],
\end{equation}
where $r_{ij}$ is the distance between particles $i$ and $j$, and $\epsilon$ sets the energy scale. The characteristic length scale $\sigma_{ij}$ is defined by the sum of the particle radii, $\sigma_{ij} = \sigma_{\mathrm{coeff}} (R_i + R_j)$, where $R_i$ and $R_j$ are the radii of the interacting particles. 

The system evolves under athermal, quasistatic conditions. We employ the Fast Inertial Relaxation Engine (FIRE) algorithm~\cite{bitzek2006structural} to minimize the total potential energy of the system. In each simulation step, the system is relaxed until the root-mean-square force on the particles, $F_{\text{rms}}$, falls below a convergence tolerance of $10^{-3}$. 

\subsection{Training Protocol}
The training process involves applying coordinated cyclic displacements to specific subsets of particles to encode a desired input--output response. We designate a fixed set of particles as ``inputs'' and a dynamically selected set of particles as ``outputs.''

\textit{Region driving.} In this work, the input particles are fixed in identity throughout a given run and are driven according to the prescribed cyclic protocol. By contrast, the output particles are identified dynamically at the start of each training cycle by a fixed-region (``region driving'') rule: we define target regions in the simulation box, and for each region the particle closest to its geometric center is designated as the output particle for that cycle. Only these region-selected particles are treated as outputs; all other grains that are not input or output particles are passive. This strategy allows the material to adapt its configuration without being constrained by the specific identity of the output grains since we are training for an associative flow pattern, i.e., should the two input particles move in a particular manner, particles in the chosen output regions should move accordingly.  

\textit{Cyclic driving.} Each training cycle consists of a loading phase and an unloading phase. During the loading phase, both the fixed input particles and the region-selected output particles are actively driven. Their positions are updated linearly in small increments over $N_{\text{steps}}$ steps from their initial positions $\mathbf{r}(0)$ to a target maximum displacement $\mathbf{r}(0) + \mathbf{A}$, where $\mathbf{A}$ is the prescribed target displacement vector. After each incremental displacement of the driven particles, they are held fixed while the remaining ``passive'' particles are allowed to relax via FIRE minimization. In the unloading phase, the driven particles are returned linearly to their cycle-start positions, again with energy minimization at each step. We will explore how the driving amplitudes for inputs and outputs can be modulated (e.g., ramped up or down) over the course of a training epoch to determine the retrieval performance. 

\textit{Intermittent relaxation.} To enhance the stability of the learned behavior, we implement an intermittent relaxation protocol. Specifically, in every block of $N_r$ training cycles, the outputs are released from their driven constraints for one cycle (the $N_r$th cycle), while the inputs continue to be driven. In these relaxation cycles, only the input particles are driven, allowing the output particles and the surrounding medium to relax naturally in response to the input forcing. This reduces residual stress near the outputs and helps prevent the system from becoming trapped in highly stressed configurations that are incompatible with the passive mechanical response required during retrieval. See Fig. 1(a) for a schematic of the training protocol.

\subsection{Retrieval and Evaluation}
During retrieval, the input particles are driven according to the same cyclic protocol used in training, while the output particles are left free to move in response to forces transmitted through the granular network. See Fig. 1(b). We quantify associative-memory performance from the output displacements measured at the point of maximum input displacement (end of the loading phase). Let $\Delta \mathbf{r}_i$ denote the displacement of the $i$th output particle relative to its position at the start of the cycle. We then define a cosine-similarity score, $S(\{\Delta \mathbf{r}_i\})$, to quantify the performance. Because different tasks encode different target relations, the form of $S$ depends on the task as follows. 

\textit{Coherence task (easy task).}
When the desired associative-memory response is that the two outputs move in the same direction, we quantify coherence by the cosine of the angle between the two output displacement directions,
\begin{equation}
S_{\mathrm{coh}} = 
\frac{\Delta \mathbf{r}_1 \cdot \Delta \mathbf{r}_2}{|\Delta \mathbf{r}_1|\,|\Delta \mathbf{r}_2|}
= \cos\theta,
\end{equation}
where $\theta$ is the angle between the two output displacement vectors at maximum input displacement. In this case, $S_{\mathrm{coh}}\approx 1$ indicates strongly aligned output motion.

\textit{Directional tasks (hard task and quadrupolar task).}
When each output has a prescribed target direction $\hat{\mathbf{u}}_i$, we define the directional score
\begin{equation}
S_{\mathrm{dir}} = \frac{1}{N_{\text{out}}} \sum_{i=1}^{N_{\text{out}}}
\frac{\Delta \mathbf{r}_i \cdot \hat{\mathbf{u}}_i}{|\Delta \mathbf{r}_i|}.
\end{equation}
Here $S_{\mathrm{dir}}\approx 1$ indicates that outputs move along the trained directions, while $S_{\mathrm{dir}}\approx 0$ implies uncorrelated motion.

In the Results section we report $S_{\mathrm{coh}}$ for the easy task and $S_{\mathrm{dir}}$ for tasks with prescribed output directions, averaging over many independent realizations. See Fig. 1(c) for a schematic of the cosine similarity performance.

\textit{Contact-change metrics and train--retrieval mismatch.}
We define a contact between particles $i$ and $j$ using the minimum-image separation under periodic boundary conditions.
Let $r_{ij}(t)$ denote this minimum-image distance evaluated at the \emph{cycle start} of cycle $t$.
The pair is deemed to be in contact if
\begin{equation}
r_{ij}(t) < r_c \equiv 2^{1/6}\,\sigma_{\mathrm{coeff}}(R_i+R_j),
\end{equation}
where $r_c$ is the location of the Lennard--Jones potential minimum.

For each cycle $t$, we record the set of contacting pairs at the cycle start,
\begin{equation}
\mathcal{C}(t)=\{(i,j): i<j \ \text{and}\ r_{ij}(t)< r_c\},
\end{equation}
and define the baseline contact set as $\mathcal{C}(0)$, taken from the first cycle start.

Equivalently, we encode contact status as a binary contact-indicator vector $\mathbf{c}(t)$ over all unordered pairs $(i,j)$ with $i<j$,
with components
\begin{equation}
c_{ij}(t)=\mathbf{1}[(i,j)\in \mathcal{C}(t)].
\end{equation}
We then define the \emph{contact-change signature} relative to baseline as the bitwise XOR
\begin{equation}
\boldsymbol{\delta}(t)=\mathbf{c}(t)\oplus \mathbf{c}(0),
\end{equation}
so that $\delta_{ij}(t)=1$ iff the contact status of pair $(i,j)$ at the cycle start differs between cycle $t$ and cycle $0$.

The \emph{total contact change} (TCC) is the normalized Hamming distance from baseline,
\begin{equation}
\mathrm{TCC}(t)=\frac{\|\boldsymbol{\delta}(t)\|_1}{|\mathcal{C}(0)|}
=\frac{\sum_{i<j}\delta_{ij}(t)}{|\mathcal{C}(0)|}.
\end{equation}

Note that $\mathrm{TCC}(t)$ is not bounded by $1$ in our definition. We count all contact-state flips relative to the baseline (including both broken baseline contacts and newly formed contacts) and normalize only by the number of baseline contacts, $|\mathcal{C}(0)|$. When many baseline contacts break while many new contacts form, the number of flips can exceed $|\mathcal{C}(0)|$, so $\mathrm{TCC}(t)$ can be larger than $1$.

To quantify train--retrieval consistency, we compare the contact networks in the final trained state
and the late-time steady state reached during retrieval.
Let $t_{\mathrm{tr}}^\ast$ denote the final training cycle,
and let $t_{\mathrm{te}}^\ast$ denote a late retrieval cycle after the contact statistics have reached a plateau.
We define the corresponding binary contact-indicator vectors
\begin{equation}
\mathbf{c}_{\mathrm{tr}} \equiv \mathbf{c}(t_{\mathrm{tr}}^\ast),
\qquad
\mathbf{c}_{\mathrm{te}} \equiv \mathbf{c}(t_{\mathrm{te}}^\ast),
\end{equation}
where $c_{ij}(t)=\mathbf{1}[(i,j)\in \mathcal{C}(t)]$ for all unordered pairs $i<j$.

The \emph{terminal contact-change signature} between the two end states is the bitwise XOR
\begin{equation}
\boldsymbol{\delta}_{\mathrm{tr}\rightarrow \mathrm{te}}
\equiv
\mathbf{c}_{\mathrm{tr}}\oplus \mathbf{c}_{\mathrm{te}},
\end{equation}
so that $\delta_{ij}=1$ if the contact status of pair $(i,j)$ differs between the trained state and the steady retrieval state.

We then define the \emph{train--retrieval mismatch} as the normalized Hamming distance between the two end states,
\begin{equation}
\mathcal{M}_{\mathrm{tr}\rightarrow \mathrm{te}}
=
\frac{\|\boldsymbol{\delta}_{\mathrm{tr}\rightarrow \mathrm{te}}\|_1}{\|\mathbf{c}_{\mathrm{tr}}\|_1},
\end{equation}
i.e., the fraction of contacts in the trained state whose contact status differs in the steady retrieval state.
Here $\|\cdot\|_1$ denotes the $\ell_1$ norm, so $\|\mathbf{c}_{\mathrm{tr}}\|_1$ is the number of contacts in the trained state. 

For later reference, we define the \emph{terminal} training-stage contact reorganization as
\begin{equation}
\mathrm{TCC}_{\mathrm{tr}} \equiv \mathrm{TCC}(t_{\mathrm{tr}}^\ast),
\end{equation}
i.e., the value of $\mathrm{TCC}(t)$ at the final training cycle $t_{\mathrm{tr}}^\ast$.

\section{Results}

\subsection{Overview of Learning Outcomes}
As in many learning systems, the distinction between ``easy'' and ``hard'' tasks is often framed in terms of task complexity, or how compatible a desired input--output mapping is with the system’s inductive bias~\cite{Bengio2013}. Tasks that align with system's preferred modes of response are typically solved with little training, whereas tasks that conflict with those modes require substantial internal reorganization. While this notion is well established in artificial intelligence, analogous ideas have recently been explored in physical learning systems with fixed architecture, where the intrinsic physical response plays a role similar to an inductive bias in shaping learnability~\cite{li2024training,stern2025physical}. We will find that a closely related notion of task complexity arises in reconfigurable particle packings. Here, the material itself possesses intrinsic collective mechanical responses that act as a form of physical inductive bias. Learning therefore depends not only on the training protocol, but on whether the desired mapping is compatible with these native modes of deformation.

To guide interpretation of the detailed case studies that follow, Table~\ref{tab:learning_summary} summarizes the qualitative learning outcomes across task types and training protocols. The table highlights three key observations. The first is that when the desired response aligns with intrinsic collective modes (low-complexity tasks), the untrained system already performs well and training provides little benefit. The second is when the desired mapping conflicts with intrinsic mechanics (high-complexity tasks), learning is limited unless training protocols reduce the mismatch between training and retrieval conditions. Finally, in some settings, training does not create a new mechanical response but instead selects between competing intrinsic responses, effectively biasing an underlying bistability. In the subsections below, we show how these outcomes emerge from the interplay between intrinsic mechanical response, training-induced rearrangements, and the degree of mismatch between training and retrieval conditions. We will start with an easy association task, followed by hard association tasks and conclude with intermediate tasks.

\begin{table*}[t]
\centering
\caption{Summary of associative-learning outcomes across task types and training protocols. The easy task measures \emph{coherence} between the two outputs using $S_{\mathrm{coh}}$, while the hard and intermediate (quadrupolar) tasks measure \emph{directional} accuracy using $S_{\mathrm{dir}}$. For each protocol we report the sustained untrained retrieval-time score $S_{\text{untrained}}$, the terminal training-stage contact reorganization $\mathrm{TCC}_{\mathrm{tr}}$ (end-of-training total contact change), the terminal train--retrieval mismatch $\mathcal{M}_{\mathrm{tr}\rightarrow \mathrm{te}}$ (fraction of trained contacts that differ in the late-time steady retrieval state; see Sec.~\ref{sec_methodology}), and the approximate performance gain $S_{\text{trained}}-S_{\text{untrained}}$ relative to the untrained baseline under the same retrieval protocol. In the intermediate rows, ``annealing $A_{\mathrm{out}}$'' means linearly ramping down the output-drive amplitude during training, ``tuning $\sigma_{\mathrm{coeff}}$'' means changing the Lennard--Jones length prefactor in $\sigma_{ij}=\sigma_{\mathrm{coeff}}(R_i+R_j)$, and ``increasing input size'' means enlarging the driven input-particle radii (input amplification). Here “$\cdots$” means “as above”: we start from the protocol in the preceding row and add the extra changes listed for this row (i.e., modifications are cumulative across these rows).}
\label{tab:learning_summary}
\begin{tabular}{l l l l l l}
\hline
Task Type & Training Protocol & $S_{\text{untrained}}$ & $\mathrm{TCC}_{\mathrm{tr}}$ & $\mathcal{M}_{\mathrm{tr}\rightarrow \mathrm{te}}$ & $S_{\text{trained}}-S_{\text{untrained}}$ \\
\hline

Easy
& Same-direction output drive 
& $S_{\mathrm{coh}}\sim 0.65$ 
& $\sim 0.3$ 
& $\sim 0.15$ 
& None ($\sim 0$) \\
%& Training perturbs favorable response

Easy
& Opposite-direction output drive 
& $S_{\mathrm{coh}}\sim 0.65$ 
& $\sim 0.6$ 
& $\sim 0.17$ 
& None ($\sim 0$) \\
%& Intrinsic response dominates 

Hard
& Opposite-direction output drive
& $S_{\mathrm{dir}}\sim 0.1$ 
& $\sim 0.65$ 
& $\sim 0.17$ 
& Low ($\sim 0.1$) \\
%& Conflict with intrinsic collective mode 

Hard
& $\cdots$ + relaxation 
& $S_{\mathrm{dir}}\sim 0.1$ 
& $\sim 1.5$ 
& $\sim 0.05$ 
& Moderate ($\sim 0.2$) \\
%& Reduced train--retrieval mismatch 

Hard
& Same-direction output drive 
& $S_{\mathrm{dir}}\sim -0.1$ 
& $\sim 0.35$ 
& $\sim 0.16$ 
& Moderate high ($\sim 0.4$) \\

Hard
& $\cdots$ + relaxation 
& $S_{\mathrm{dir}}\sim -0.1$ 
& $\sim 0.65$ 
& $\sim 0$ 
& High ($\sim 0.6$) \\
%& Training selects for one intrinsic drift 

Intermediate
& Quadrupolar drive 
& $S_{\mathrm{dir}}\sim 0.4$ 
& $\sim 0.4$ 
& $\sim 0.18$ 
& None ($\sim 0$) \\

Intermediate
& $\cdots$ + relaxation 
& $S_{\mathrm{dir}}\sim 0.4$ 
& $\sim 1.1$ 
& $\sim 0.01$ 
& Low ($\sim 0.1$) \\
%& Geometry stabilizes learned mapping 

Intermediate
& $\cdots$ + annealing $A_{\mathrm{out}}$ + tuning $\sigma_{\mathrm{coeff}}$
& $S_{\mathrm{dir}}\sim 0.4$ 
& $\sim 1.0$ 
& $\sim 0$ 
& Moderate ($\sim 0.2$) \\

Intermediate
& $\cdots$ + increasing input size
& $S_{\mathrm{dir}}\sim 0.4$ 
& $\sim 1.2$ 
& $\sim 0.01$ 
& Moderate high ($\sim 0.4$) \\

\hline
\end{tabular}
\end{table*}

\subsection{Easy task}
\begin{figure*}[h]
	\centering
	\includegraphics[width=0.9\textwidth]{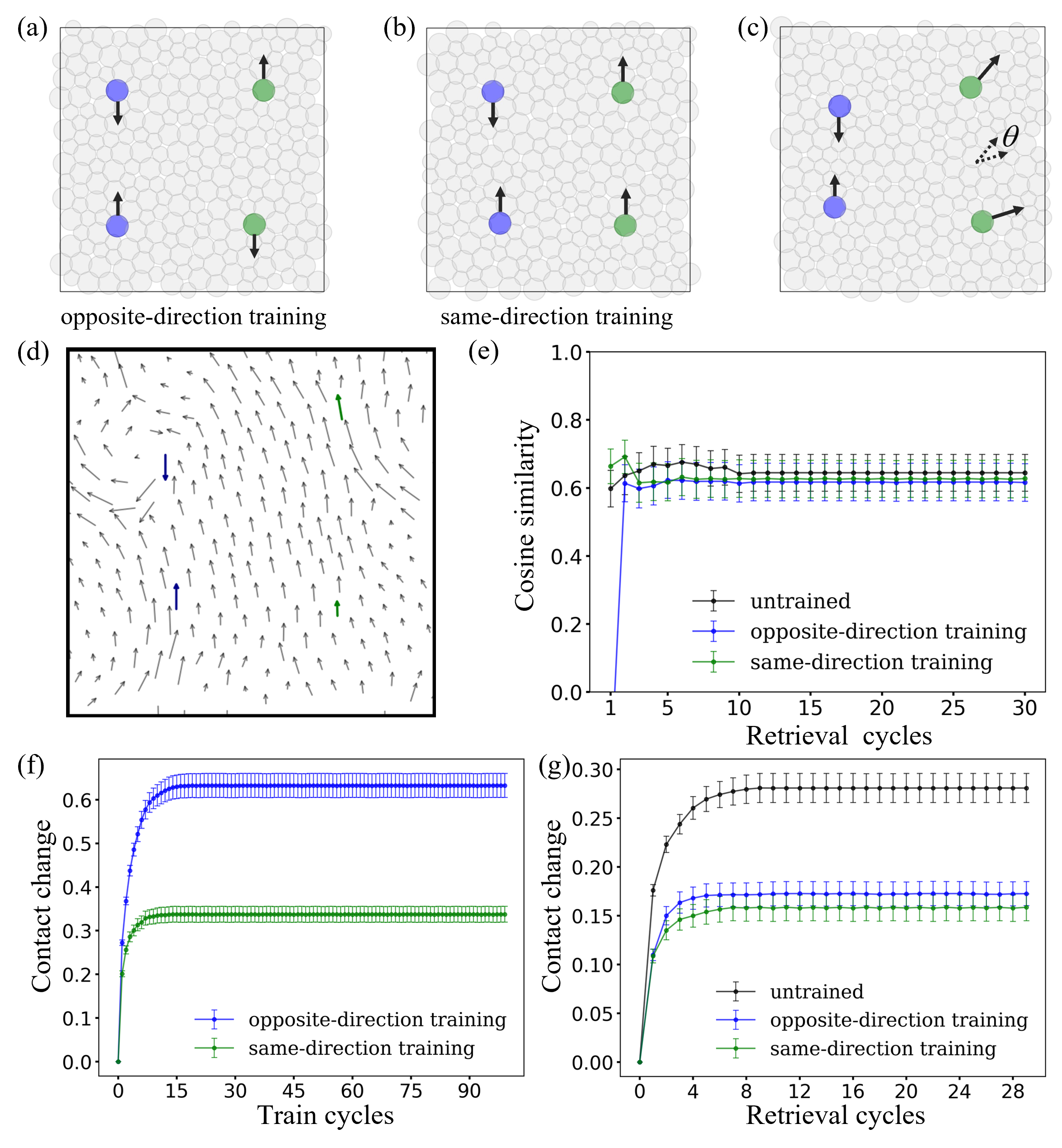}
	\caption{{\it Easy task.} 
(a) Opposite-direction training schematic: two fixed input particles (blue) are cyclically driven while two region-selected output particles (green) are driven in opposite directions (arrows show imposed displacements). 
(b) Same-direction training schematic: same inputs, but the two outputs are driven in the same direction during training. 
(c) Retrieval schematic: only the inputs are driven; at maximum input displacement we measure the instantaneous output displacement directions and define the angle $\theta$ between them. 
(d) Representative displacement field at maximum input displacement during retrieval for the untrained system (blue: inputs; green: outputs; gray: passive particles). 
(e) Cosine similarity versus retrieval-cycle number, computed as $\cos\theta$: black, untrained (direct retrieval); blue, opposite-direction training; green, same-direction training. 
(f) Total contact change $\mathrm{TCC}(t)$ measured at the cycle start during training, defined relative to the baseline contact network in the first cycle (see text for definition).
(g) Total contact change $\mathrm{TCC}(t)$ measured at the cycle start during retrieval, again referenced to the same baseline contact network $\mathbf{C}_0$ from the first cycle.}

\label{fig:easytask}
\end{figure*}

We begin by examining a task in which the desired response is largely consistent with the material's intrinsic mechanical response. In this \emph{easy task}, the goal is not to enforce a particular \emph{absolute} direction of motion for each output particle, but rather to promote \emph{coherence} between the two outputs in the associative-memory response. Concretely, during retrieval we quantify whether the two outputs move approximately parallel to one another in any direction when only the inputs are driven.

We consider two training protocols that use the same fixed input particles but different output-driving patterns. In \emph{same-direction training} [Fig.~\ref{fig:easytask}(b)], both region-selected outputs are driven in the same direction (upward) during training. In \emph{opposite-direction training} [Fig.~\ref{fig:easytask}(a)], the two region-selected outputs are driven in opposite directions (one upward and one downward) while the inputs are driven according to their prescribed cyclic motion. This second type of training will test whether coherence reflects a true mechanical bias or merely an incidental feature of the untrained response since the training protocol that actively opposes it. In other words, this opposite-driving protocol serves as a minimal test of whether training can override intrinsic mechanical tendencies and induce a qualitatively different response. In both cases, the output particles are selected anew at the start of each cycle by the region-driving rule described in Sec.~\ref{sec_methodology}, whereas the input particles remain fixed in identity throughout. 

After training, we evaluate associative-memory behavior by driving only the two inputs and leaving all other particles (including the outputs) undriven [Fig.~\ref{fig:easytask}(c)]. At the moment of maximum input displacement (half-cycle), we record the instantaneous displacement directions of the two output particles and define $\theta$ as the angle between these two directions. Because the easy task targets \emph{alignment} of the two outputs, the relevant performance measure -- the cosine similarity plotted in Fig.~\ref{fig:easytask}(e) is computed from $\cos\theta$, so that $\cos\theta\simeq 1$ indicates strongly aligned output motion, while $\cos\theta\simeq 0$ corresponds to orthogonal responses.

The key physical reason this task is ``easy'' is revealed by the retrieving the  displacement field for the untrained system shown in Fig.~\ref{fig:easytask}(d). Driving the two left inputs tends to expel nearby particles laterally, and the right half of the system responds with a collective drift: the particles on the right preferentially move \emph{together} either upward or downward during a retrieval cycle. Importantly, although the \emph{sign} of this drift (up versus down) fluctuates between realizations, the \emph{coherence} of the motion implies that the two outputs typically share nearly the same displacement direction in a given retrieval. As a result, even without any training, the untrained system already exhibits a high cosine similarity, as shown by the black curve in Fig.~\ref{fig:easytask}(e).

Interestingly, applying training does not improve performance for this easy task; instead, both same-direction training and opposite-direction training lead to a modest reduction in cosine similarity (green and blue curves in Fig.~\ref{fig:easytask}(e)). For same-direction training, the cosine similarity in the \emph{first} retrieval cycle is slightly higher than the untrained baseline (green curve), consistent with the fact that during training the two outputs are driven in the same direction. However, this bias is transient: already by the third retrieval cycle the cosine similarity drops to slightly below the black curve, indicating that the training-induced preference does not persist once the outputs are released and the system relaxes under input-only driving. For opposite-direction training, the cosine similarity in the \emph{first} retrieval cycle is negative on average (blue curve), consistent with the outputs being driven in opposite directions during training, but this anti-aligned response is likewise short-lived and rebounds toward the untrained baseline by the second retrieval cycle. In both cases, the long-cycle cosine similarity settles to a similar level. This train--retrieval contrast is visualized in Fig.~\ref{fig:easyDisplacementField_S1}: during training the output directions are externally imposed [panels (a,c)], whereas during retrieval, the undriven outputs relax into the collective displacement field [panels (b,d)], explaining why the cycle-1 bias in Fig.~\ref{fig:easytask}(e) is quickly erased.

The contact-network diagnostics support this interpretation. Figure~\ref{fig:easytask}(f) shows the total contact change $\mathrm{TCC}(t)$ during training, computed from the contact mask at the cycle start relative to the baseline network in the first cycle. Opposite-direction training (blue) produces a larger $\mathrm{TCC}$ and reaches its plateau more slowly than same-direction training (green), indicating more extensive and longer-lived rearrangements are required to accommodate the anti-aligned output driving, which is less compatible with the intrinsic same-direction drift of the right-hand region (Fig.~\ref{fig:easytask}(d)). 

Crucially, Fig.~\ref{fig:easytask}(g) shows that upon switching to retrieval  (input-only driving), both trained systems undergo rapid further reorganization: $\mathrm{TCC}$ rises within the first few retrieval cycles to $\gtrsim 0.15$, implying that a substantial fraction of baseline contacts have changed relative to the pre-training reference. This early contact reshuffling coincides with the quick loss of the cycle-1 bias in Fig.~\ref{fig:easytask}(e), consistent with the view that the output-driven training imprints a short-lived structural bias that is promptly erased when the outputs are released. Moreover, the untrained system (black) continues to evolve over more retrieval cycles before stabilizing, whereas the trained systems reach their retrieval $\mathrm{TCC}$ plateau earlier; this accelerated approach reflects training-induced pre-rearrangement of the packing and helps explain the slight downward shift of the long-cycle cosine similarity relative to the untrained baseline. Overall, actively driving the outputs during training perturbs an otherwise favorable intrinsic response: it induces contact-network reorganization that, once retrieval begins, relaxes and reshapes the subsequent drift, leading to a small net degradation in the steady associative-memory score (see the training-stage displacement fields in Fig.~\ref{fig:easyDisplacementField_S1}(a,c), where the imposed output drives visibly distort the local response near the outputs relative to the undriven field in Fig.~\ref{fig:easytask}(d)).

\subsection{Hard task}

\begin{figure*}[h]
	\centering
	\includegraphics[width=0.9\textwidth]{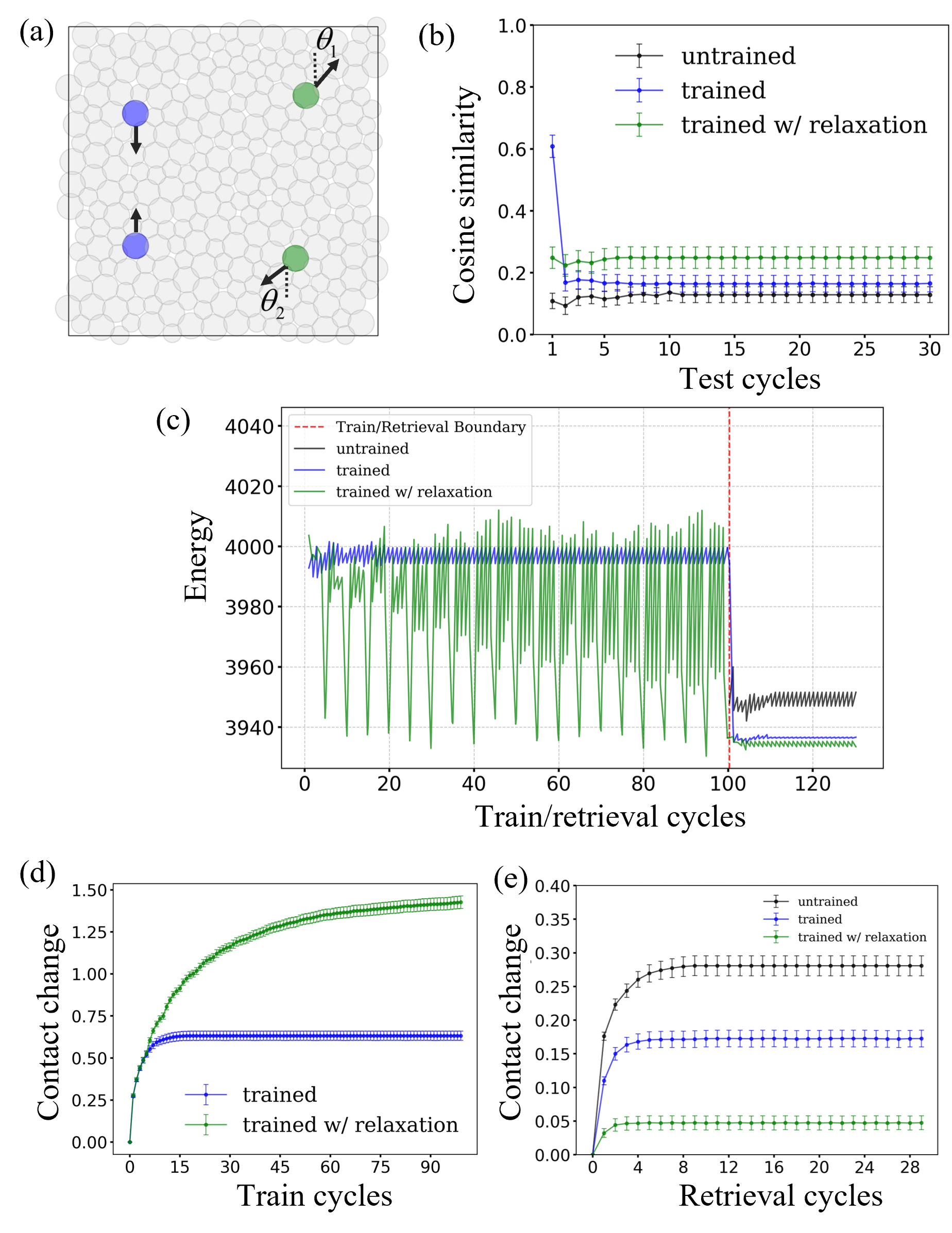}
	\caption{{\it Hard task (opposite-direction training: opposite-direction output drive)} (a) Retrieval schematic: only input particles are driven; snapshot at the halfway point of a retrieval cycle when the input particle reaches its maximum displacement; $\theta_1$ and $\theta_2$ are measured as the deviation angles between output-particle displacement directions and their target displacement directions. (b) Cosine similarity vs test cycles, computed from $\theta_1$ and $\theta_2$. Black curve shows the untrained case (no training, direct retrieval). Blue curve uses region driving rule. Green curve uses the same protocol with an output-relaxation step during training, in which every block of five cycles leaves the outputs undriven in the fifth cycle to allow relaxation. (c) Average energy versus cycle number. Each curve represents the mean across 100 independent runs. The vertical red dashed line separates the training phase (left) from the retrieval phase (right), with connecting lines showing the transition between phases. Data points correspond to the energy at the initial position and maximum displacement position within each cycle. (d) Total contact change, $\mathrm{TCC}(t)$, versus training cycle number for the trained protocols (blue/green), computed from the contact status at the cycle start relative to the baseline contact set at the first cycle start. (e) $\mathrm{TCC}(t)$ versus retrieval cycle number (same definition as in (d)). Error bars indicate variability across 100 independent realizations.}
	\label{fig:type1a}
\end{figure*}

Having established in the easy task that the right half of the packing tends to respond with a \emph{coherent} drift (upward or downward) under input-only driving [Fig.~\ref{fig:easytask}(d)], we now formulate a mapping that is \emph{incompatible} with this dominant collective mode. Specifically, we consider a hard task in which the two outputs are required to move in \emph{opposite} directions during retrieval. As in Sec.~\ref{sec_methodology}, the input particles are fixed in identity and driven cyclically, while the outputs are selected by region driving during training and left free during retrieval. The retrieval geometry and the deviation angles used to construct the cosine similarity are shown schematically in Fig.~\ref{fig:type1a}(a).

The difficulty of this task is immediately reflected in the untrained performance. The black curve in Fig.~\ref{fig:type1a}(b) remains low across retrieval cycles, indicating that without training the outputs rarely realize the required anti-aligned motion. This poor baseline follows naturally from the coherent right-side drift observed in Fig.~\ref{fig:easytask}(d): when the right half moves largely as a block, both outputs tend to be advected in the \emph{same} direction (either up or down), whereas the hard-task target demands that one output move up while the other moves down. Thus the intrinsic mechanical response that benefited the easy task now directly obstructs learning, making this mapping a stringent test for our local training protocol.

Applying \emph{opposite-direction training} (opposite-direction output drive) yields only a modest improvement (blue curve in Fig.~\ref{fig:type1a}(b)), and the cosine similarity remains well below unity. Moreover, the trained curve exhibits a pronounced drop after the first retrieval cycle: the cosine similarity starts relatively high in the first cycle but quickly falls to values comparable to the untrained baseline. This transient arises from the mismatch between training and retrieval constraints. During training, the region-selected outputs are actively driven every cycle and therefore do not fully relax; once retrieval begins, the outputs are undriven and undergo relaxation away from the driven configuration, degrading the apparent associative-memory performance.

To reduce this train--retrival mismatch, we incorporate intermittent relaxation during training, as described in Sec.~\ref{sec_methodology}. In every $N_r$-th training cycle (here $N_r=5$), the outputs are left undriven while the inputs continue to be driven, allowing the output neighborhood to relax under input forcing. The choice $N_r=5$ is guided by a sweep over relaxation intervals shown in Supplemental Fig.~\ref{fig:hardtask1relaxation}. This modification increases the cosine similarity (green curve in Fig.~\ref{fig:type1a}(b)) and suppresses the sharp post-transition drop seen without relaxation. The energetic signature is consistent with this interpretation: without relaxation, the energy quickly plateaus during training and then drops abruptly when switching to retrieval (blue curve in Fig.~\ref{fig:type1a}(c)), consistent with a sudden relaxation of the previously constrained outputs. With intermittent relaxation, the energy decays more gradually during training and exhibits a smaller drop at the onset of retrieval (green curve in Fig.~\ref{fig:type1a}(c)), correlating with improved and more stable associative-memory performance.

The contact-change metric $\mathrm{TCC}(t)$ [Fig.~\ref{fig:type1a}(d,e)] provides a complementary, structural view consistent with the energy trends in Fig.~\ref{fig:type1a}(c). During training [Fig.~\ref{fig:type1a}(d)], the no-relaxation protocol (blue) rapidly reaches a plateau in $\mathrm{TCC}(t)$, indicating that the cycle-start contact network quickly becomes stationary relative to the baseline, whereas with intermittent relaxation (green) $\mathrm{TCC}(t)$ keeps increasing over many cycles before leveling off, signaling continued contact rearrangements and ongoing learning. In retrieval  [Fig.~\ref{fig:type1a}(e)], the untrained case shows the largest mismatch (near $\sim 30\%$), and training without relaxation reduces it only partially (still $\gtrsim 15\%$), while intermittent relaxation suppresses the mismatch to a few percent (about $\sim 5\%$). This improved retention of the trained contact network under input-only driving explains why the relaxation protocol yields a more stable retrieval response and higher cosine similarity than the no-relaxation protocol in Fig.~\ref{fig:type1a}(b). 

\begin{figure*}[h]
	\centering
	\includegraphics[width=0.9\textwidth]{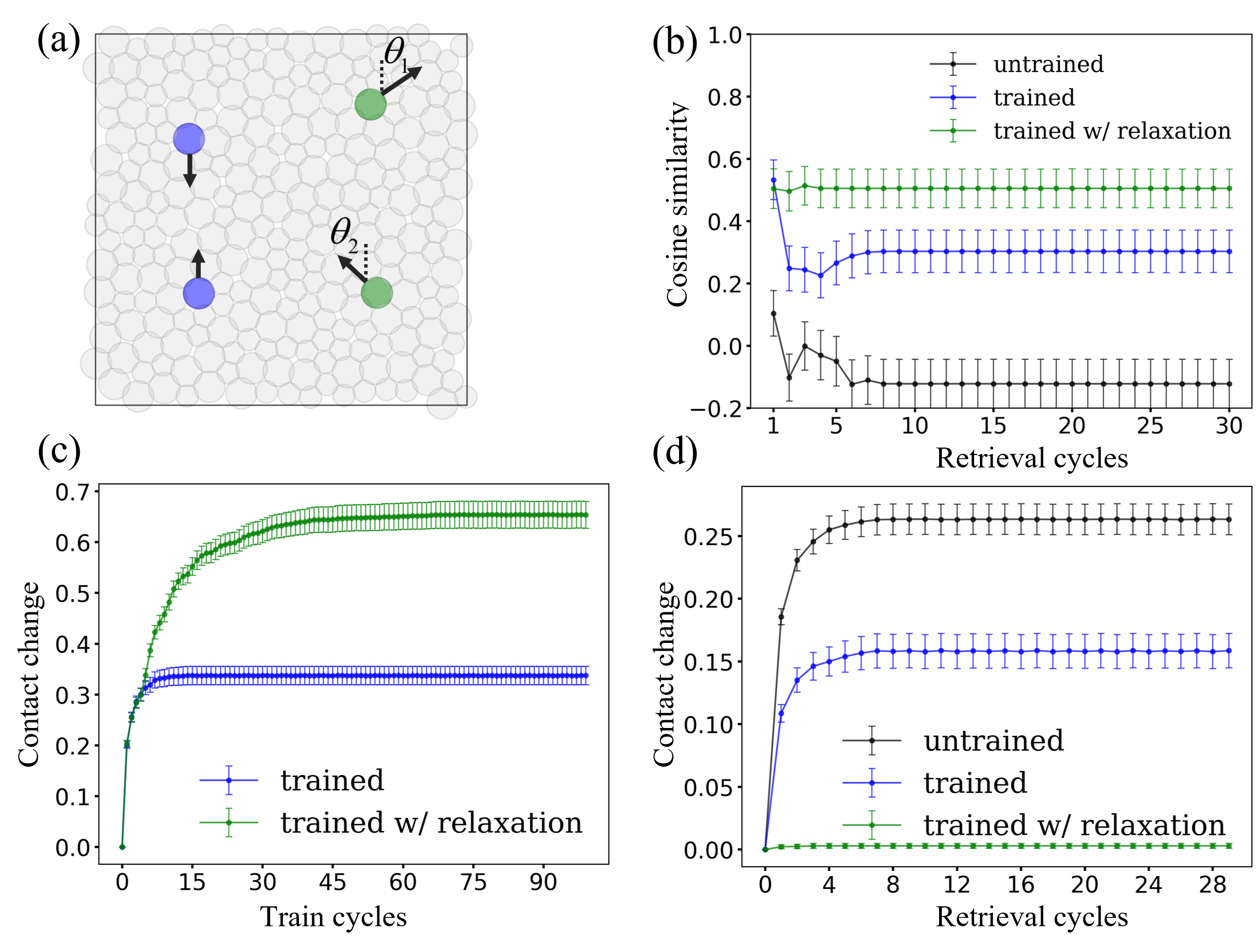}
	\caption{{\it Hard task (same-direction training: same-direction output drive)} (a) Retrieval schematic: only input particles are driven; snapshot at the halfway point of a retrieval cycle when the input particle reaches its maximum displacement; $\theta_1$ and $\theta_2$ are measured as the deviation angles between output-particle displacement directions and their target displacement directions. (b) Cosine similarity vs retrieval cycles, computed from $\theta_1$ and $\theta_2$. Black curve shows the untrained case (no training, direct retrieval). Blue curve uses region driving of the outputs in every training cycle. Green curve uses the same protocol with an output-relaxation step during training, in which every block of five cycles leaves the outputs undriven in the fifth cycle to allow relaxation. (c) Total contact change, $\mathrm{TCC}(t)$, versus training cycle number for the trained protocols (blue/green), computed from the contact status at the cycle start relative to the baseline contact set at the first cycle start. (d) $\mathrm{TCC}(t)$ versus retrieval cycle number (same definition as in (c)). Error bars indicate variability across 100 independent realizations. }
	\label{fig:type1b}
\end{figure*}

We also examine a related hard setting using \emph{same-direction training} (same-direction output drive), in which during training both outputs are driven upward, and the retrieval objective is that both outputs move upward under input-only driving. The retrieval schematic is shown in Fig.~\ref{fig:type1b}(a). Despite the change in target, the untrained cosine similarity remains low (black curve in Fig.~\ref{fig:type1b}(b)), because the intrinsic response on the right side still has two competing collective outcomes (upward or downward drift), and only a subset of realizations are consistent with an ``upward'' target. In this case, however, training is substantially more effective: the trained response (blue curve) is markedly higher than the untrained baseline, and adding intermittent relaxation further boosts performance (green curve), consistent with the idea that training can bias the system toward selecting the desired branch of the otherwise bistable collective drift.

This interpretation is supported by the contact-network evolution quantified by the total contact change $\mathrm{TCC}(t)$ in Fig.~\ref{fig:type1b}(c,d). During training, $\mathrm{TCC}(t)$ continues to grow for substantially longer and reaches a larger plateau when intermittent relaxation is included (green curve in Fig.~\ref{fig:type1b}(c)) than in the baseline protocol (blue curve), indicating that relaxation sustains additional particle rearrangements as the system explores and stabilizes a modified contact network consistent with the intended upward-response branch. During retrieval, the contrast is even sharper: with relaxation, $\mathrm{TCC}(t)$ drops to essentially zero (green curve in Fig.~\ref{fig:type1b}(d)), implying that the trained contact network is preserved under input-only driving and that train--test mismatch is strongly suppressed. Without relaxation, the trained system still exhibits substantial contact reorganization during the first few test cycles ($\mathrm{TCC}\gtrsim 15\%$, blue curve), indicating a pronounced mismatch and partial loss of the training-imposed structure once the outputs are released. This reduction of mismatch explains why the relaxation protocol avoids the first-cycle degradation and yields a higher, more stable cosine similarity than the baseline trained case in Fig.~\ref{fig:type1b}(b).

\begin{figure*}[h]
	\centering
	\includegraphics[width=0.9\textwidth]{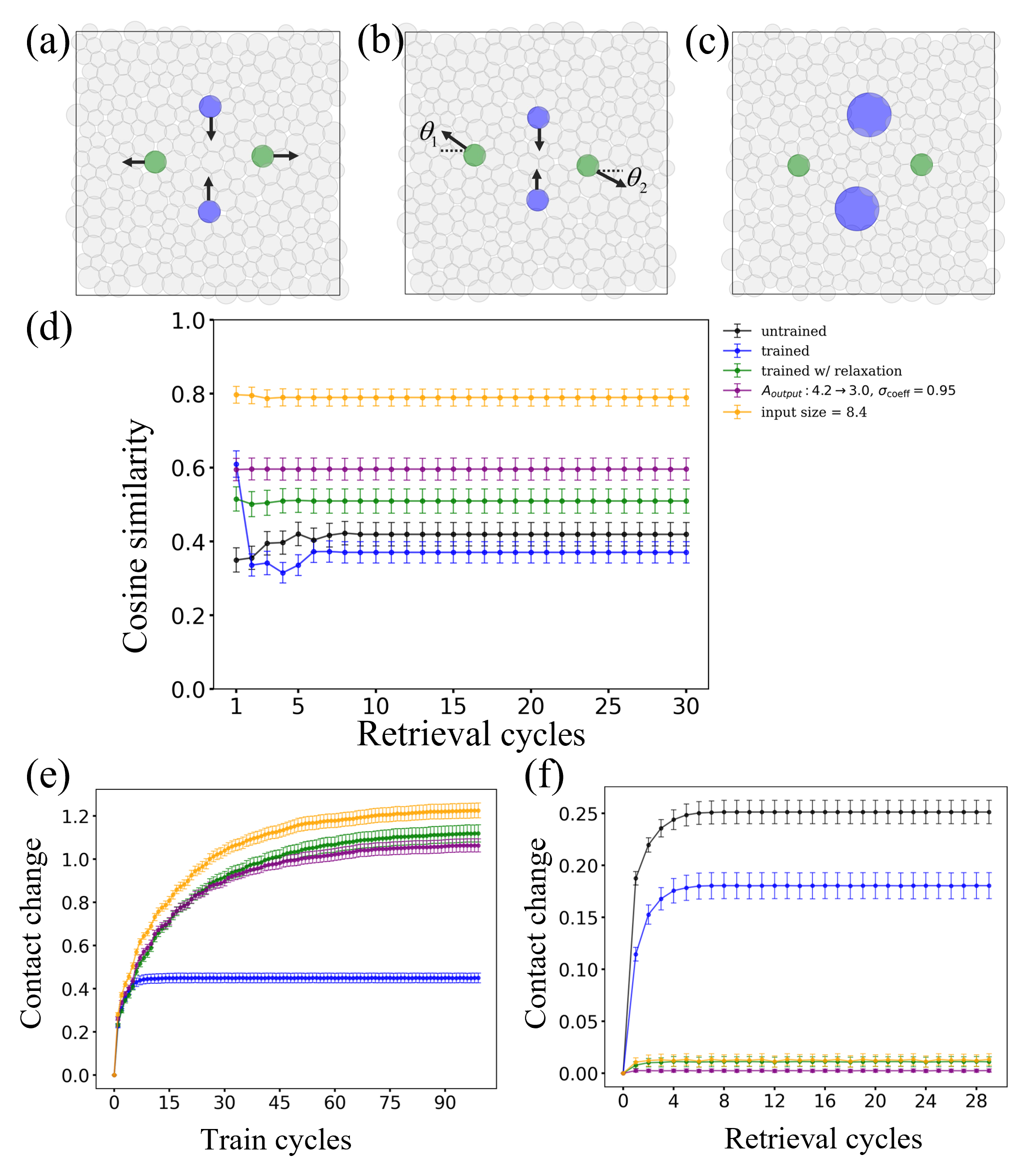}
	\caption{{\it Quadrupolar task.} (a) Training protocol: two input particles (blue) are cyclically driven in opposite vertical directions, while two output particles (green) are cyclically driven in opposite horizontal directions (arrows indicate the imposed driving directions), forming a quadrupolar pattern. (b) Retrieval protocol: only the input particles are driven; output particles are undriven and respond mechanically. At maximum input displacement we measure the output displacement directions and define deviation angles $\theta_1$ and $\theta_2$ relative to the target directions (dotted arrows). (c) Same geometry as (a) but with enlarged input particles (used in the parameter variations reported in panel (d)). (d) Cosine similarity versus retrieval cycle number for the quadrupolar task. Black: untrained (direct retrieval without training). Blue: baseline training. Green: training with intermittent relaxation (every 5 training cycles the outputs are not driven). Purple: same as green but with an \emph{annealed} output drive during training, where the output amplitude is ramped down linearly from $A_{\mathrm{out}}=4.2$ (first training cycle) to $A_{\mathrm{out}}=3.0$ (final training cycle), together with a reduced Lennard--Jones length prefactor ($\sigma_{\mathrm{coeff}}=0.95$ instead of $1.0$), which shortens the interaction length scale.
 Yellow: same as purple with increased input particle size (input size $=8.4$). (e) Total contact change $\mathrm{TCC}(t)$ versus training cycle number for the same protocols/labels as in (d).
(f) $\mathrm{TCC}(t)$ versus retrieval cycle number for the same protocols/labels as in (d) (see text for definition of $\mathrm{TCC}$).Error bars indicate variability across 100 independent realizations. }
	\label{fig:type3}
\end{figure*}

\subsection{Quadrupolar task}
We first consider an \emph{intermediate} setting in which the input--output geometry is rearranged into a quadrupolar pattern, with two inputs placed near the vertical midline and two outputs placed near the horizontal midline [Fig.~\ref{fig:type3}(a,b)]. In contrast to the easy and hard tasks, where the dominant response of the right half of the packing is a collective vertical drift, the quadrupolar layout is designed to couple the vertical input forcing to a transverse (horizontal) response at the output locations. During training, the two fixed input particles are driven cyclically in opposite vertical directions, while the two region-selected output particles are driven cyclically in opposite horizontal directions [Fig.~\ref{fig:type3}(a)]. During retrieval, only the inputs are driven and the outputs respond mechanically [Fig.~\ref{fig:type3}(b)]. We quantify associative-memory performance using the directional score $S_{\mathrm{dir}}$ defined in Sec.~\ref{sec_methodology}, computed from the deviation angles $\theta_1$ and $\theta_2$ measured at maximum input displacement. As seen in the untrained displacement field (Supplemental Fig.~\ref{fig:quadrupoleDisplacementField_S3}(a)), the intrinsic response under input-only driving can organize into a side-wide up/down drift rather than the intended transverse quadrupolar flow, motivating this as an \emph{intermediate} (not intrinsically easy) task.

The black curve in Fig.~\ref{fig:type3}(d) shows the untrained response, with a sustained similarity $S_{\mathrm{dir}}\simeq 0.40$. Applying the baseline training protocol (blue curve) does not improve performance: $S_{\mathrm{dir}}$ settles near $\simeq 0.37$ and exhibits the same characteristic train--retrieval degradation seen in the hard task. This poor retention is reflected structurally by the retrieval-stage contact mismatch: the untrained system stabilizes at $\mathrm{TCC}\approx 25\%$ [Fig.~\ref{fig:type3}(f), black], while the baseline-trained system still shows $\mathrm{TCC}\approx 18\%$ [blue], indicating substantial reorganization once the outputs are released and explaining why the trained response does not persist. This is also evident in the displacement fields: the final training-cycle snapshot shows the imposed quadrupolar driving organizing local flow (Supplemental Fig.~\ref{fig:quadrupoleDisplacementField_S3}(b)), but the corresponding late-time retrieval snapshot quickly deviates once outputs are released (Supplemental Fig.~\ref{fig:quadrupoleDisplacementField_S3}(c)), consistent with rapid forgetting.

Introducing intermittent relaxation during training substantially improves performance. In the relaxation protocol (green curve), every fifth training cycle leaves the outputs undriven while continuing to drive the inputs, aligning training constraints more closely with retrieval. This raises the sustained similarity to $S_{\mathrm{dir}}\simeq 0.50$ [Fig.~\ref{fig:type3}(d), green] and strongly suppresses retrieval-time mismatch: $\mathrm{TCC}$ drops from the $\sim 15$--$25\%$ range down to $\sim 1\%$ [Fig.~\ref{fig:type3}(f), green]. Consistent with this reduced mismatch, the training- and retrieval-stage displacement fields become much more similar when relaxation is included (Supplemental Fig.~\ref{fig:quadrupoleDisplacementField_S3}(d,e)). At the same time, the training-stage contact evolution reaches a large but steady plateau of $\mathrm{TCC}\approx 110\%$ [Fig.~\ref{fig:type3}(e), green], indicating substantial training-induced reorganization of the contact topology. 

With this relaxation protocol as a baseline, we next ask which additional knobs can further improve $S_{\mathrm{dir}}$ \emph{without reintroducing} train--retrieval mismatch. The parameter sweep in Fig.~\ref{fig:quadrupole_parameters} addresses this question by varying the output-drive amplitude $A_{\mathrm{out}}$ and the interaction length prefactor $\sigma_{\mathrm{coeff}}$, and tracking both performance [Fig.~\ref{fig:quadrupole_parameters}(a)] and contact reorganization [Fig.~\ref{fig:quadrupole_parameters}(b,c)]. Two robust trends emerge. First, reducing $A_{\mathrm{out}}$ decreases training-stage contact change, as expected: smaller imposed output displacements induce fewer rearrangements. Second, decreasing $\sigma_{\mathrm{coeff}}$ from $1.00$ to $0.95$ or $0.90$ also tends to \emph{lower} the training-stage $\mathrm{TCC}$, consistent with the idea that shortening the effective interaction range weakens force transmission from the driven outputs into the surrounding network, thereby reducing the number of contact-status changes even if the effective free volume increases. Importantly, for most parameter sets the reduction in training-stage $\mathrm{TCC}$ is modest, so the learned response is not strongly limited by insufficient rearrangements. Instead, the dominant discriminator is the \emph{retrieval-stage} mismatch in Fig.~\ref{fig:quadrupole_parameters}(c): at $\sigma_{\mathrm{coeff}}=1.00$ or $0.90$, lowering $A_{\mathrm{out}}$ does not strongly suppress mismatch, whereas at the intermediate value $\sigma_{\mathrm{coeff}}=0.95$ it can reduce the retrieval-stage $\mathrm{TCC}$ to below $1\%$. This identifies $\sigma_{\mathrm{coeff}}=0.95$ as a practical sweet spot: it is low enough to reduce train--retrieval mismatch when $A_{\mathrm{out}}$ is annealed, yet not so low that diminished interaction range starves the system of the rearrangements needed to encode a robust quadrupolar response.

Motivated by these trends, we anneal the output driving amplitude during training from $A_{\mathrm{out}}=4.2$ to $A_{\mathrm{out}}=3.0$, and simultaneously set $\sigma_{\mathrm{coeff}}=0.95$ (purple curve). Compared to relaxation alone, this protocol increases the sustained similarity from $\simeq 0.50$ to $S_{\mathrm{dir}}\simeq 0.60$ [Fig.~\ref{fig:type3}(d), purple]. Figure~\ref{fig:type3}(e) shows that the associated training-stage rearrangement activity is \emph{slightly reduced} : the plateau decreases from $\sim 110\%$ (green) to $\sim 105\%$ (purple). The performance gain instead correlates with a cleaner handoff to retrieval: the retrieval-stage mismatch is reduced from $\sim 1\%$ (green) to essentially $0$--$0.5\%$ (purple) [Fig.~\ref{fig:type3}(f)], indicating that the learned contact topology is more faithfully preserved under input-only driving. Thus, in this regime the primary benefit of lowering $\sigma_{\mathrm{coeff}}$ and annealing $A_{\mathrm{out}}$ is not to increase training-stage plasticity, but to \emph{stabilize} the trained configuration against test-time relaxation, consistent with the tighter agreement between the training- and retrieval-stage displacement fields (Supplemental Fig.~\ref{fig:quadrupoleDisplacementField_S3}(f,g)).

Finally, we strengthen the \emph{input} signal by increasing the input particle size to $8.4$ (yellow curve; schematic in Fig.~\ref{fig:type3}(c)). This modification yields the highest performance, with $S_{\mathrm{dir}}\simeq 0.80$ [Fig.~\ref{fig:type3}(d), yellow]. Input amplification markedly increases training-stage reorganization: $\mathrm{TCC}$ rises to $\sim 120\%$ [Fig.~\ref{fig:type3}(e), yellow], indicating substantially more extensive rearrangements during learning, and it also visibly amplifies the induced flow field (Supplemental Fig.~\ref{fig:quadrupoleDisplacementField_S3}(h,i)). Strikingly, this larger training-stage plasticity does \emph{not} come at the cost of additional mismatch: the retrieval-stage $\mathrm{TCC}$ remains at $\sim 1\%$, comparable to the relaxation-only protocol [Fig.~\ref{fig:type3}(f), yellow vs green]. The combination of high training-stage reorganization with low retrieval-stage mismatch explains the large boost in $S_{\mathrm{dir}}$.

\subsection{Learning with rearrangements}

Learning association formation in particle packings is fundamentally shaped by contact rearrangements, which introduce discrete structural changes into an otherwise continuous mechanical response. Under cyclic driving, grains repeatedly form and break contacts, producing history-dependent evolution of rigidity/force networks. These rearrangements act as state transitions between mechanically stable configurations, so training is not merely a smooth tuning of parameters but a navigation through a high-dimensional landscape punctuated by topological changes in the contact network. Given the Lennard--Jones interactions implemented here, we define contacts using the criterion in Sec.~\ref{sec_methodology} (pairs on the repulsive side of the LJ minimum). Because stress transmission pathways depend sensitively on the contact network, each rearrangement can redirect how input forces propagate to distant regions, effectively rewriting the material’s internal “wiring diagram.” Learning therefore emerges from the selective stabilization of rearrangement pathways that consistently produce the desired output response under repeated input cycling; importantly, these rearrangements represent \emph{discontinuous} changes in the contact network.

Learning in Artificial Intelligence is often introduced as smooth optimization in a high-dimensional parameter space, yet many modern systems operate in fundamentally non-smooth regimes~\cite{Goodfellow2016,LeCun2015,Montufar2014}. Discontinuities arise from thresholds, switching rules, and discrete decisions—such as in spiking or binary neurons, max-pooling, or decision trees—so that small parameter changes can trigger abrupt shifts in behavior~\cite{Breiman1984,Bengio2013ST,Neftci2019}. Rather than gentle basins, the objective landscape resembles a rugged terrain of plateaus, ridges, and metastable regions. Because true gradients may vanish or be undefined in non-smooth systems \cite{Bengio1994,Goodfellow2016}, practical methods introduce auxiliary smooth structure to guide learning. Surrogate or “straight-through” estimators treat hard switches as differentiable during backpropagation \cite{Bengio2013ST,Neftci2019}. Other approaches embed discontinuous choices in probabilistic or thermally smoothed descriptions, optimizing ensemble-averaged objectives or annealing softmax functions toward sharp decisions~\cite{Jang2017,Maddison2017,Hinton2015}. Even without explicit smoothing, piecewise-linear models such as ReLU networks admit a geometric interpretation: the system is linear within regions defined by fixed activation patterns, and training becomes motion through a combinatorial tiling of such regions, with sudden transitions when boundaries are crossed~\cite{Goodfellow2016,Montufar2014,Raghu2017}. From a physics viewpoint, learning in these architectures is therefore less like continuous gradient flow in a smooth potential and more like navigating a landscape of metastable configurations connected by discrete switches.

Here, we build on the notion of learning trajectories through a combinatorial tiling of linear regions, rather than descent on a single smooth manifold as is the case for smooth parameter optimization on a fixed architecture. In particle packings, contact creation and destruction generate discontinuous changes in the mechanical network, so that infinitesimal variations of the driving protocol may move the system between distinct response manifolds. More precisely, for a fixed contact network $\mathcal{C}$, small perturbations about a mechanical equilibrium produce an approximately linear response,
\begin{equation}
\Delta \mathbf{r}_{\mathrm{out}}
\approx
\mathbf{G}(\mathcal{C})\,
\Delta \mathbf{r}_{\mathrm{in}},
\end{equation}
where $\mathbf{G}(\mathcal{C})$ is an effective Green's function determined by the current contact topology and stiffness structure. A rearrangement event therefore constitutes a boundary crossing $\mathcal{C}\rightarrow \mathcal{C}'$, which instantaneously replaces one linear map with another. From this viewpoint, rearrangements are not accidents of the dynamics; they are the mechanism by which the system rewires the computation it implements.

Because the output response depends on which contact networks are visited, successful learning requires \emph{two} ingredients: (i) sufficient rearrangements during training to access advantageous effective response maps, and (ii) \emph{train--retrieval consistency}, so that retrieval revisits (or remains near) the trained contact topology under input-only driving. We quantify training-stage rearrangement activity using the total contact change $\mathrm{TCC}_{\mathrm{tr}}(t)$ (Sec.~\ref{sec_methodology}), and we quantify \emph{trained-memory association} using the terminal train--retrieval mismatch $\mathcal{M}_{\mathrm{tr}\rightarrow \mathrm{te}}$ (Sec.~\ref{sec_methodology}), defined by directly comparing the \emph{final trained} contact network to the \emph{late-time steady} contact network reached during retrieval. Importantly, $\mathrm{TCC}_{\mathrm{tr}}(t)$ curves alone do \emph{not} determine $\mathcal{M}_{\mathrm{tr}\rightarrow \mathrm{te}}$: $\mathrm{TCC}_{\mathrm{tr}}(t)$ tracks how far the system moves relative to the \emph{initial} system state before training, whereas $\mathcal{M}_{\mathrm{tr}\rightarrow \mathrm{te}}$ depends on the specific contacts that differ between the two terminal states.

\begin{figure*}[h]
	\centering
	\includegraphics[width=0.8\textwidth]{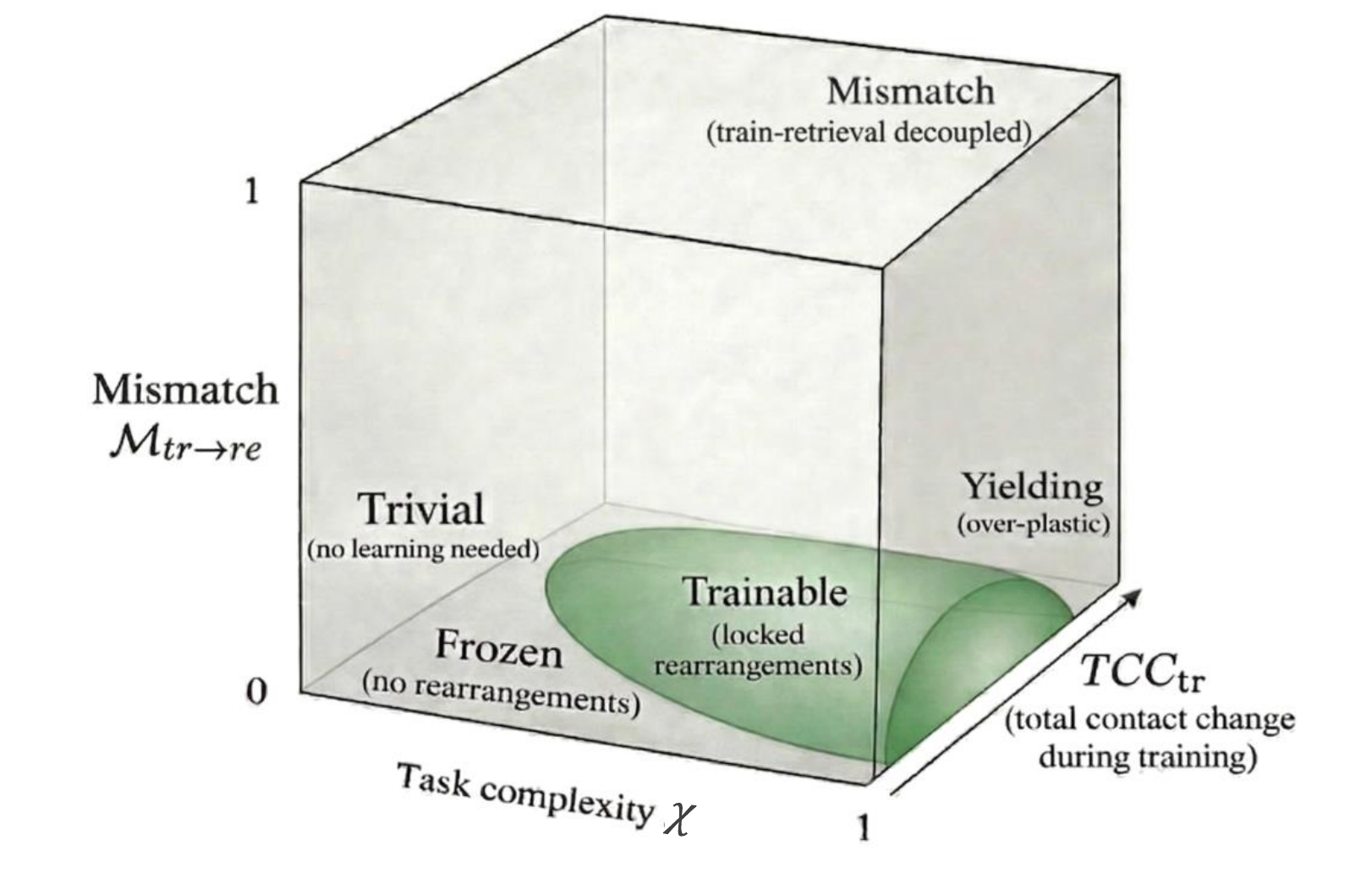}
	\caption{{\it A conceptual associative learning phase diagram for disordered particle packings.}
Given our results, we organize learning outcomes using three quantities: task complexity $\chi \equiv 1-S_{\rm untrained}$ (horizontal axis), training-stage rearrangement activity $\mathrm{TCC}_{\rm tr}$ (depth axis; total contact change during training), and the terminal train--retrieval mismatch $\mathcal{M}_{\mathrm{tr}\rightarrow \mathrm{te}}$ (vertical axis; Sec.~\ref{sec_methodology}), defined by directly comparing the final trained contact network to the late-time steady contact network in retrieval.
The green region indicates a \emph{trainable} regime, where rearrangements occur during training and are subsequently \emph{locked in} (low mismatch) so that learned behavior generalizes under input-only driving.
When $\chi\!\approx\!0$ the task is \emph{trivial} (no learning needed), when $\mathrm{TCC}_{\rm tr}\!\approx\!0$ the system is \emph{frozen} (no rearrangements), when $\mathrm{TCC}_{\rm tr}$ is too large the system is \emph{yielding} (over-plastic), and when $\mathcal{M}_{\mathrm{tr}\rightarrow \mathrm{te}}$ is large the dynamics are \emph{mismatched} (train--retrieval decoupled) and association fails.}
	\label{fig:phase_diagram}
\end{figure*}

With such a construction, we can now qualitatively discuss a learning phase diagram organized by three quantities (Fig.~\ref{fig:phase_diagram}): (1) task complexity, $\chi \equiv 1 - S_{\mathrm{untrained}}$, with $\chi=1$ representing a highly nontrivial mapping and $\chi=0$ a response that is already physically natural; (2) training-stage rearrangement activity, quantified by the terminal total contact change $\mathrm{TCC}_{\mathrm{tr}}$; and (3) terminal train--retrieval mismatch, $\mathcal{M}_{\mathrm{tr}\rightarrow \mathrm{te}}$ (both defined in Sec.~\ref{sec_methodology}). 
We label a protocol as \emph{trainable} only if training produces a measurable improvement in the retrieval-time score relative to the untrained baseline, i.e., $S_{\mathrm{trained}}>S_{\mathrm{untrained}}$ for the same task and retrieval protocol. Within this space, we identify four broad regimes. In the {\bf intrinsic} regime ($\chi\approx 0$), the target response is already aligned with the material’s native mechanics, so learning is unnecessary and training can even degrade performance (as in the easy task). In the {\bf frozen} regime ($\mathrm{TCC}_{\mathrm{tr}}\approx 0$), rearrangements are too rare or absent, preventing the system from switching to new effective response maps. In the {\bf learnable} regime, rearrangements during training are substantial ($\mathrm{TCC}_{\mathrm{tr}}>0$) yet become \emph{locked in} so that $\mathcal{M}_{\mathrm{tr}\rightarrow \mathrm{te}}$ remains small; in this regime training yields $S_{\mathrm{trained}}>S_{\mathrm{untrained}}$ and the learned behavior is retained under input-only driving (as in the best quadrupolar protocols). Finally, in the {\bf yielding} (over-plastic) regime, $\mathrm{TCC}_{\mathrm{tr}}$ is so large that the contact topology does not stabilize; as a result $\mathcal{M}_{\mathrm{tr}\rightarrow \mathrm{te}}$ remains large and trajectories fail to repeat, so retrieval-time retention breaks down despite strong plastic activity.

The easy task provides a clean illustration of the {\bf intrinsic} regime, where $\chi\approx 0$ and learning is unnecessary. In this setting, input-only driving already excites a dominant coherent drift mode on the right [Fig.~\ref{fig:easytask}(d)], so the untrained system exhibits high baseline performance [Fig.~\ref{fig:easytask}(e), black] without any programming. Actively driving the outputs during training perturbs this favorable basin and induces additional rearrangements [Fig.~\ref{fig:easytask}(f)], but the perturbation does not survive the transition to input-only driving: the trained responses lose their cycle-1 bias within the first few retrieval cycles [Fig.~\ref{fig:easytask}(e)], coincident with rapid contact reshuffling during retrieval [Fig.~\ref{fig:easytask}(g)]. Thus, even though training generates nonzero rearrangement activity, it effectively increases train--retrieval mismatch and slightly degrades the steady performance relative to the intrinsic response, consistent with the phase-diagram expectation that when $\chi\simeq 0$ training can be neutral or even counterproductive.

For fixed task complexity $\chi$, increasing training-stage rearrangements from $\mathrm{TCC}_{\mathrm{tr}}\approx 0$ initially moves the system out of the frozen regime by enabling exploration of new contact networks; learning is optimized in an intermediate window where rearrangements are sufficient to discover useful response maps but can still be retained so that $\mathcal{M}_{\mathrm{tr}\rightarrow \mathrm{te}}$ stays small (Fig.~\ref{fig:phase_diagram}). This trend is borne out directly by the quadrupolar task in Fig.~\ref{fig:type3}(d--f): the baseline training protocol (blue) is effectively non-trainable, with $S_{\mathrm{dir}}\approx 0.37$ falling \emph{below} the untrained baseline $S_{\mathrm{dir}}\approx 0.40$ (black), and it is also the protocol with the weakest training-stage reorganization in Fig.~\ref{fig:type3}(e), consistent with being closer to the frozen side of the phase diagram. In contrast, protocols that increase training-stage contact reorganization shift the system into the trainable window: intermittent relaxation (green) raises the training-stage contact change to $\mathrm{TCC}_{\mathrm{tr}}\approx 110\%$ and boosts performance to $S_{\mathrm{dir}}\approx 0.50$, while the combined protocol (output-amplitude annealing $A_{\mathrm{out}}:4.2\!\rightarrow\!3.0$ with $\sigma_{\mathrm{coeff}}=0.95$, purple) maintains substantial rearrangements ($\mathrm{TCC}_{\mathrm{tr}}\approx 105\%$--$110\%$) and further improves to $S_{\mathrm{dir}}\approx 0.60$; amplifying the inputs (yellow) pushes training-stage rearrangements to $\mathrm{TCC}_{\mathrm{tr}}\approx 120\%$ and yields the highest score $S_{\mathrm{dir}}\approx 0.80$ [Fig.~\ref{fig:type3}(d,e)]. Thus, within the quadrupolar task at fixed $\chi$, increasing $\mathrm{TCC}_{\mathrm{tr}}$ correlates with crossing from a frozen/non-trainable response (blue, below black) into a regime where learning surpasses the untrained baseline (green/purple/yellow), consistent with the phase-diagram picture that nonzero training plasticity is necessary for programmability. Beyond this window, however, further increases in rearrangement activity would be expected to enter an over-plastic yielding regime in which contact topology fails to stabilize and mismatch grows, suppressing generalization (Fig.~\ref{fig:phase_diagram}).

Taken together, the quadrupolar and hard tasks directly demonstrate that {\bf reducing mismatch} is the key control knob for accessing the trainable region in Fig.~\ref{fig:phase_diagram}. In the quadrupolar geometry, the protocols that achieve the highest $S_{\mathrm{dir}}$ are precisely those that nearly eliminate retrieval-stage reorganization: the retrieval-stage contact-change curves saturate at $\sim 0\%$--$1\%$ [Fig.~\ref{fig:type3}(f)], indicating that the trained contact topology is retained under input-only driving and hence that the terminal mismatch $\mathcal{M}_{\mathrm{tr}\rightarrow \mathrm{te}}$ is small. In the hard tasks, where the target mapping conflicts with the dominant collective drift, learning remains fragile unless this same train--retrieval drift is actively controlled: intermittent relaxation during training suppresses retrieval-stage contact reorganization to $\sim 5\%$ for opposite-direction training (vs.\ $\gtrsim 15\%$ without relaxation and $\sim 30\%$ untrained) [Fig.~\ref{fig:type1a}(e)], and to nearly $0\%$ for same-direction training [Fig.~\ref{fig:type1b}(d)], in parallel with the improved and more stable cosine-similarity curves [Fig.~\ref{fig:type1a}(b); Fig.~\ref{fig:type1b}(b)].

Varying the training protocol provides a practical means of navigating the phase diagram by trading off exploration against retention. Increasing the relaxation frequency (equivalently, introducing more frequent opportunities for the system to settle toward nearby mechanical equilibria) tends to suppress train--retrieval drift and thus reduce $\mathcal{M}_{\mathrm{tr}\rightarrow \mathrm{te}}$ (e.g., the relaxation protocols in the hard and quadrupolar tasks; Supplemental Fig.~\ref{fig:hardtask1relaxation}). Likewise, annealing $A_{\mathrm{out}}$ and tuning $\sigma_{\mathrm{coeff}}$ can reduce mismatch without strongly starving training of rearrangements (Supplemental Fig.~\ref{fig:quadrupole_parameters}). Finally, increasing the input-particle size amplifies the imposed input signal and can increase training-stage rearrangements while still maintaining low mismatch (Fig.~\ref{fig:type3}). In this sense, intermittent relaxation, output-amplitude annealing, modest tuning of $\sigma_{\mathrm{coeff}}$, and input amplification act primarily by enabling sufficient exploration while keeping $\mathcal{M}_{\mathrm{tr}\rightarrow \mathrm{te}}$ small (improving retention), thereby converting plastic but non-generalizing dynamics into trainable behavior.

Recent work has established a geometric picture of how memory can emerge during cyclic optimization in disordered systems through Gradient Discontinuity Learning (GDL)~\cite{zu2025learning}. In that framework, contact changes generate discontinuities in the gradient of the trained quantity with respect to control parameters. When these discontinuities are of the appropriate type, ascent and descent trajectories become trapped along them, progressively funneling the dynamics toward a marginally absorbing manifold (MAM), and return--point memory follows from the stability of this manifold under cyclic driving. Our observations are broadly consistent with this picture—rearrangements are indispensable, cyclic protocols stabilize special trajectories, and reproducibility improves as training proceeds—but the emphasis here is different. Rather than focusing on how discontinuities guide motion through parameter space, we focus on the learning consequences of the contact networks that are visited. For a fixed contact network, the mechanical response is approximately linear, so each accessible network corresponds to a distinct effective input--output operator; rearrangements therefore act as switches between operators. From this perspective, learning requires both exploration (rearrangements) and retention (low $\mathcal{M}_{\mathrm{tr}\rightarrow \mathrm{te}}$): rearrangements enlarge the repertoire of available operators, while successful training demands that retrieval remain within the same operator sequence discovered during training.

\section{Discussion}

We have demonstrated that a reconfigurable particle packing subjected to purely local cyclic driving can be trained to exhibit associative-memory-like behavior with emergent weight updates. Unlike prior realizations of mechanical allostery and programmable metamaterials, which rely on modifying effective couplings on a fixed underlying architecture, our system continually rearranges its contact network during training. Despite this structural plasticity, we find that stable input–output associations can be learned, stored, and retrieved, revealing that reconfiguration does not preclude memory. Rather, it can serve as a physical mechanism for modifying effective interactions between distant regions of a material. This perspective connects naturally to recent work that frames learning in reconfigurable materials as a feedback-modified aging process, in which repeated driving and response-dependent adjustments progressively bias a disordered configuration toward functionally useful basins~\cite{anisetti2023emergent}.

A central result of our study is that associative learning in reconfigurable matter is not uniform across tasks. Instead, we identify three qualitatively distinct regimes—easy, hard, and intermediate—whose existence reflects the compatibility between the desired input–output mapping and the system’s intrinsic mechanical response. In easy tasks, the untrained material already exhibits collective motion aligned with the target association, so our training protocols confer little advantage and may even degrade performance by perturbing an otherwise favorable response. In hard tasks, the target association conflicts with dominant mechanical modes, resulting in fragile and weak learning. Between these extremes, we find an intermediate regime in which appropriate geometric placement of inputs and outputs can stabilize reproducible learned behavior. This classification highlights a fundamental constraint on learning in physical systems: not all mappings are equally realizable, and learnability is governed by the mechanical susceptibility of the medium itself.

This perspective extends and complements prior work on mechanical allostery. In fixed-architecture networks, learning is achieved by sculpting stress-propagation pathways through changes in spring constants, rest lengths, or similar continuous parameters. Here, by contrast, learning emerges through the reorganization of the contact network, which reshapes how stresses are transmitted and redistributed. In this sense, structural plasticity plays a role analogous to adaptive weights in neural networks, modifying effective couplings without explicit parameter tuning. These results differ from other physical learning research going beyond fixed architecture networks, such as the cytoskeleton with severing, but with an explicit weight update rule~\cite{banerjee2026learning}.
With emergent weight updates, rather than tuning responses on a static graph, one can reshape the graph itself. The quadrupolar geometry illustrates this principle. In this intermediate regime, the geometric arrangement of inputs and outputs suppresses dominant collective drift modes and instead promotes transverse responses that can be robustly stabilized by training. This finding emphasizes that learning in reconfigurable systems is inherently geometric: the spatial embedding of inputs and outputs determines which mechanical modes are accessible and which can be selectively reinforced. From this viewpoint, training does not merely adjust a response—it navigates a constrained landscape of mechanically admissible mappings.

A further key observation is the importance of train–retrieval consistency (or mismatch). When outputs are actively driven during training but left passive during retrieval, the system can relax into configurations that are incompatible with the learned response. Introducing intermittent relaxation cycles mitigates this mismatch by allowing the material to explore configurations consistent with retrieval-time constraints. This echoes ideas from both machine learning and materials science: generalization requires that training conditions reflect deployment conditions. In physical systems, this constraint manifests mechanically, through residual stresses and relaxation pathways.

Although we have framed our results in terms of associative memory, they point toward a broader implication: memory in reconfigurable matter may be fundamentally state-based rather than parameter-based. Disordered solids respond to driving through localized, history-dependent rearrangements, which endow the system with intrinsic memory of past deformations. When viewed abstractly, the mechanically stable configurations of such a system and the transitions between them form a state-transition structure reminiscent of a finite state machine. While we do not explicitly identify or manipulate this structure here, our results suggest that local cyclic driving can sculpt reproducible paths through this state space, effectively stabilizing particular input–output relations. This raises a natural question for future work: can the state-transition structure itself be trained? In other words, can one systematically create, annihilate, or rewire the localized rearrangements that define the material’s internal memory, thereby modifying the transition graph directly? Such a capability would constitute learning at the level of a finite state machine, rather than merely learning a response on a fixed graph. Our results provide a necessary first step toward this goal by demonstrating that reconfigurable matter can support stable associations at all.

Several limitations of our present study point to fruitful directions for further exploration. First, we have focused on two-dimensional, frictionless packings under quasistatic driving. Extending these ideas to frictional packings~\cite{henkes2016rigid,liu2021spongelike} or active packings~\cite{henkes2011active} may reveal qualitatively new learning regimes. Second, exploring more complex or hierarchical driving schemes may allow multiple associations to be stored simultaneously, raising questions about capacity, interference, and forgetting \cite{Keim2019memory}. Finally, while we characterize learnability phenomenologically, a more microscopic understanding of which rearrangements encode memory and associations would enable a more predictive theory of learning in disordered solids.

Taken together, our results suggest that reconfigurable matter occupies a distinct and largely unexplored regime of physical learning. Unlike fixed-architecture networks, where learning consists of tuning parameters on a static graph, reconfigurable systems learn by reshaping their own internal organization. This structural plasticity imposes constraints on what can be learned, but also enables new forms of adaptation. By identifying regimes of learnability and demonstrating associative memory in a particle packing, we establish a minimal framework for studying learning in disordered, reconfigurable media. We anticipate that this framework will prove useful for understanding a  broad class of particle-based systems whose function emerges from the interplay of rigidity, rearrangement, and history dependence. 

Finally, just as Feynman argued that to understand something one must build it, we suggest that to understand adaptive materials we must be able to train them.
In this spirit, we train a particle packing so that localized rearrangements emerge in ways that support a desired input–output response, rather than simply waiting for such rearrangements to emerge spontaneously and attempting to retrodict their occurrence. Simply put, we train for emergence.

\section{Acknowledgements}
Tao Zhang acknowledges financial support from the NSFC/China via award 22303051. The computations in this paper were run on the $\pi$ 2.0 and the Siyuan-1 cluster supported by the Center for High Performance Computing at Shanghai Jiao Tong University. JMS acknowledges financial support from the National Science Foundation via DMR-2204312.

\bibliography{manuscript}

%merlin.mbs apsrev4-1.bst 2010-07-25 4.21a (PWD, AO, DPC) hacked
%Control: key (0)
%Control: author (8) initials jnrlst
%Control: editor formatted (1) identically to author
%Control: production of article title (-1) disabled
%Control: page (0) single
%Control: year (1) truncated
%Control: production of eprint (0) enabled
\begin{thebibliography}{52}%
\makeatletter
\providecommand \@ifxundefined [1]{%
 \@ifx{#1\undefined}
}%
\providecommand \@ifnum [1]{%
 \ifnum #1\expandafter \@firstoftwo
 \else \expandafter \@secondoftwo
 \fi
}%
\providecommand \@ifx [1]{%
 \ifx #1\expandafter \@firstoftwo
 \else \expandafter \@secondoftwo
 \fi
}%
\providecommand \natexlab [1]{#1}%
\providecommand \enquote  [1]{``#1''}%
\providecommand \bibnamefont  [1]{#1}%
\providecommand \bibfnamefont [1]{#1}%
\providecommand \citenamefont [1]{#1}%
\providecommand \href@noop [0]{\@secondoftwo}%
\providecommand \href [0]{\begingroup \@sanitize@url \@href}%
\providecommand \@href[1]{\@@startlink{#1}\@@href}%
\providecommand \@@href[1]{\endgroup#1\@@endlink}%
\providecommand \@sanitize@url [0]{\catcode `\\12\catcode `\$12\catcode
  `\&12\catcode `\#12\catcode `\^12\catcode `\_12\catcode `\%12\relax}%
\providecommand \@@startlink[1]{}%
\providecommand \@@endlink[0]{}%
\providecommand \url  [0]{\begingroup\@sanitize@url \@url }%
\providecommand \@url [1]{\endgroup\@href {#1}{\urlprefix }}%
\providecommand \urlprefix  [0]{URL }%
\providecommand \Eprint [0]{\href }%
\providecommand \doibase [0]{http://dx.doi.org/}%
\providecommand \selectlanguage [0]{\@gobble}%
\providecommand \bibinfo  [0]{\@secondoftwo}%
\providecommand \bibfield  [0]{\@secondoftwo}%
\providecommand \translation [1]{[#1]}%
\providecommand \BibitemOpen [0]{}%
\providecommand \bibitemStop [0]{}%
\providecommand \bibitemNoStop [0]{.\EOS\space}%
\providecommand \EOS [0]{\spacefactor3000\relax}%
\providecommand \BibitemShut  [1]{\csname bibitem#1\endcsname}%
\let\auto@bib@innerbib\@empty
%</preamble>
\bibitem [{\citenamefont {Tirion}(1996)}]{Tirion1996}%
  \BibitemOpen
  \bibfield  {author} {\bibinfo {author} {\bibfnamefont {M.~M.}\ \bibnamefont
  {Tirion}},\ }\href@noop {} {\bibfield  {journal} {\bibinfo  {journal}
  {Physical Review Letters}\ }\textbf {\bibinfo {volume} {77}},\ \bibinfo
  {pages} {1905} (\bibinfo {year} {1996})}\BibitemShut {NoStop}%
\bibitem [{\citenamefont {Atilgan}\ \emph {et~al.}(2001)\citenamefont
  {Atilgan}, \citenamefont {Durell}, \citenamefont {Jernigan}, \citenamefont
  {Demirel}, \citenamefont {Keskin},\ and\ \citenamefont
  {Bahar}}]{Atilgan2001}%
  \BibitemOpen
  \bibfield  {author} {\bibinfo {author} {\bibfnamefont {A.~R.}\ \bibnamefont
  {Atilgan}}, \bibinfo {author} {\bibfnamefont {S.~R.}\ \bibnamefont {Durell}},
  \bibinfo {author} {\bibfnamefont {R.~L.}\ \bibnamefont {Jernigan}}, \bibinfo
  {author} {\bibfnamefont {M.~C.}\ \bibnamefont {Demirel}}, \bibinfo {author}
  {\bibfnamefont {O.}~\bibnamefont {Keskin}}, \ and\ \bibinfo {author}
  {\bibfnamefont {I.}~\bibnamefont {Bahar}},\ }\href {\doibase
  10.1016/S0006-3495(01)76033-X} {\bibfield  {journal} {\bibinfo  {journal}
  {Biophysical Journal}\ }\textbf {\bibinfo {volume} {80}},\ \bibinfo {pages}
  {505} (\bibinfo {year} {2001})}\BibitemShut {NoStop}%
\bibitem [{\citenamefont {Gordon}\ \emph {et~al.}(2015)\citenamefont {Gordon},
  \citenamefont {Zimmerman}, \citenamefont {He}, \citenamefont {Miles},
  \citenamefont {Huang}, \citenamefont {Tiyanont}, \citenamefont {McArthur},
  \citenamefont {Aster}, \citenamefont {Perrimon}, \citenamefont {Loparo},\
  and\ \citenamefont {Blacklow}}]{Gordon2015Notch}%
  \BibitemOpen
  \bibfield  {author} {\bibinfo {author} {\bibfnamefont {W.~R.}\ \bibnamefont
  {Gordon}}, \bibinfo {author} {\bibfnamefont {B.}~\bibnamefont {Zimmerman}},
  \bibinfo {author} {\bibfnamefont {L.}~\bibnamefont {He}}, \bibinfo {author}
  {\bibfnamefont {L.~J.}\ \bibnamefont {Miles}}, \bibinfo {author}
  {\bibfnamefont {J.}~\bibnamefont {Huang}}, \bibinfo {author} {\bibfnamefont
  {K.}~\bibnamefont {Tiyanont}}, \bibinfo {author} {\bibfnamefont {D.~G.}\
  \bibnamefont {McArthur}}, \bibinfo {author} {\bibfnamefont {J.~C.}\
  \bibnamefont {Aster}}, \bibinfo {author} {\bibfnamefont {N.}~\bibnamefont
  {Perrimon}}, \bibinfo {author} {\bibfnamefont {J.~J.}\ \bibnamefont
  {Loparo}}, \ and\ \bibinfo {author} {\bibfnamefont {S.~C.}\ \bibnamefont
  {Blacklow}},\ }\href@noop {} {\bibfield  {journal} {\bibinfo  {journal}
  {Developmental Cell}\ }\textbf {\bibinfo {volume} {33}},\ \bibinfo {pages}
  {729} (\bibinfo {year} {2015})}\BibitemShut {NoStop}%
\bibitem [{\citenamefont {Rocks}\ \emph {et~al.}(2017)\citenamefont {Rocks},
  \citenamefont {Pashine}, \citenamefont {Bischofberger}, \citenamefont
  {Goodrich}, \citenamefont {Liu},\ and\ \citenamefont
  {Nagel}}]{rocks2017designing}%
  \BibitemOpen
  \bibfield  {author} {\bibinfo {author} {\bibfnamefont {J.~W.}\ \bibnamefont
  {Rocks}}, \bibinfo {author} {\bibfnamefont {N.}~\bibnamefont {Pashine}},
  \bibinfo {author} {\bibfnamefont {I.}~\bibnamefont {Bischofberger}}, \bibinfo
  {author} {\bibfnamefont {C.~P.}\ \bibnamefont {Goodrich}}, \bibinfo {author}
  {\bibfnamefont {A.~J.}\ \bibnamefont {Liu}}, \ and\ \bibinfo {author}
  {\bibfnamefont {S.~R.}\ \bibnamefont {Nagel}},\ }\href@noop {} {\bibfield
  {journal} {\bibinfo  {journal} {Proceedings of the National Academy of
  Sciences}\ }\textbf {\bibinfo {volume} {114}},\ \bibinfo {pages} {2520}
  (\bibinfo {year} {2017})}\BibitemShut {NoStop}%
\bibitem [{\citenamefont {Yan}\ \emph {et~al.}(2017)\citenamefont {Yan},
  \citenamefont {Ravasio}, \citenamefont {Brito},\ and\ \citenamefont
  {Wyart}}]{Yan2017PNAS}%
  \BibitemOpen
  \bibfield  {author} {\bibinfo {author} {\bibfnamefont {L.}~\bibnamefont
  {Yan}}, \bibinfo {author} {\bibfnamefont {R.}~\bibnamefont {Ravasio}},
  \bibinfo {author} {\bibfnamefont {C.}~\bibnamefont {Brito}}, \ and\ \bibinfo
  {author} {\bibfnamefont {M.}~\bibnamefont {Wyart}},\ }\href {\doibase
  10.1073/pnas.1615536114} {\bibfield  {journal} {\bibinfo  {journal}
  {Proceedings of the National Academy of Sciences}\ }\textbf {\bibinfo
  {volume} {114}},\ \bibinfo {pages} {2526} (\bibinfo {year}
  {2017})}\BibitemShut {NoStop}%
\bibitem [{\citenamefont {Coulais}\ \emph {et~al.}(2018)\citenamefont
  {Coulais}, \citenamefont {Teomy}, \citenamefont {de~Reus}, \citenamefont
  {Shokef},\ and\ \citenamefont {van Hecke}}]{Coulais2018MultiStep}%
  \BibitemOpen
  \bibfield  {author} {\bibinfo {author} {\bibfnamefont {C.}~\bibnamefont
  {Coulais}}, \bibinfo {author} {\bibfnamefont {E.}~\bibnamefont {Teomy}},
  \bibinfo {author} {\bibfnamefont {K.}~\bibnamefont {de~Reus}}, \bibinfo
  {author} {\bibfnamefont {Y.}~\bibnamefont {Shokef}}, \ and\ \bibinfo {author}
  {\bibfnamefont {M.}~\bibnamefont {van Hecke}},\ }\href {\doibase
  10.1038/nature18972} {\bibfield  {journal} {\bibinfo  {journal} {Nature}\
  }\textbf {\bibinfo {volume} {535}},\ \bibinfo {pages} {529} (\bibinfo {year}
  {2018})}\BibitemShut {NoStop}%
\bibitem [{\citenamefont {Pashine}\ \emph {et~al.}(2019)\citenamefont
  {Pashine}, \citenamefont {Hexner}, \citenamefont {Liu},\ and\ \citenamefont
  {Nagel}}]{pashine2019directed}%
  \BibitemOpen
  \bibfield  {author} {\bibinfo {author} {\bibfnamefont {N.}~\bibnamefont
  {Pashine}}, \bibinfo {author} {\bibfnamefont {D.}~\bibnamefont {Hexner}},
  \bibinfo {author} {\bibfnamefont {A.~J.}\ \bibnamefont {Liu}}, \ and\
  \bibinfo {author} {\bibfnamefont {S.~R.}\ \bibnamefont {Nagel}},\ }\href@noop
  {} {\bibfield  {journal} {\bibinfo  {journal} {Science Advances}\ }\textbf
  {\bibinfo {volume} {5}},\ \bibinfo {pages} {eaax4215} (\bibinfo {year}
  {2019})}\BibitemShut {NoStop}%
\bibitem [{\citenamefont {Hexner}\ \emph {et~al.}(2020)\citenamefont {Hexner},
  \citenamefont {Pashine}, \citenamefont {Liu},\ and\ \citenamefont
  {Nagel}}]{hexner2020effect}%
  \BibitemOpen
  \bibfield  {author} {\bibinfo {author} {\bibfnamefont {D.}~\bibnamefont
  {Hexner}}, \bibinfo {author} {\bibfnamefont {N.}~\bibnamefont {Pashine}},
  \bibinfo {author} {\bibfnamefont {A.~J.}\ \bibnamefont {Liu}}, \ and\
  \bibinfo {author} {\bibfnamefont {S.~R.}\ \bibnamefont {Nagel}},\ }\href@noop
  {} {\bibfield  {journal} {\bibinfo  {journal} {Physical Review Research}\
  }\textbf {\bibinfo {volume} {2}},\ \bibinfo {pages} {043231} (\bibinfo {year}
  {2020})}\BibitemShut {NoStop}%
\bibitem [{\citenamefont {Scellier}\ and\ \citenamefont
  {Bengio}(2017)}]{scellier2017equilibrium}%
  \BibitemOpen
  \bibfield  {author} {\bibinfo {author} {\bibfnamefont {B.}~\bibnamefont
  {Scellier}}\ and\ \bibinfo {author} {\bibfnamefont {Y.}~\bibnamefont
  {Bengio}},\ }\href@noop {} {\bibfield  {journal} {\bibinfo  {journal}
  {Frontiers in computational neuroscience}\ }\textbf {\bibinfo {volume}
  {11}},\ \bibinfo {pages} {24} (\bibinfo {year} {2017})}\BibitemShut {NoStop}%
\bibitem [{\citenamefont {Stern}\ \emph {et~al.}(2021)\citenamefont {Stern},
  \citenamefont {Hexner}, \citenamefont {Rocks},\ and\ \citenamefont
  {Liu}}]{stern2021supervised}%
  \BibitemOpen
  \bibfield  {author} {\bibinfo {author} {\bibfnamefont {M.}~\bibnamefont
  {Stern}}, \bibinfo {author} {\bibfnamefont {D.}~\bibnamefont {Hexner}},
  \bibinfo {author} {\bibfnamefont {J.~W.}\ \bibnamefont {Rocks}}, \ and\
  \bibinfo {author} {\bibfnamefont {A.~J.}\ \bibnamefont {Liu}},\ }\href@noop
  {} {\bibfield  {journal} {\bibinfo  {journal} {Physical Review X}\ }\textbf
  {\bibinfo {volume} {11}},\ \bibinfo {pages} {021045} (\bibinfo {year}
  {2021})}\BibitemShut {NoStop}%
\bibitem [{\citenamefont {Anisetti}\ \emph
  {et~al.}(2023{\natexlab{a}})\citenamefont {Anisetti}, \citenamefont
  {Scellier},\ and\ \citenamefont {Schwarz}}]{anisetti2023}%
  \BibitemOpen
  \bibfield  {author} {\bibinfo {author} {\bibfnamefont {V.~R.}\ \bibnamefont
  {Anisetti}}, \bibinfo {author} {\bibfnamefont {B.}~\bibnamefont {Scellier}},
  \ and\ \bibinfo {author} {\bibfnamefont {J.~M.}\ \bibnamefont {Schwarz}},\
  }\href@noop {} {\bibfield  {journal} {\bibinfo  {journal} {Phys. Rev.
  Research}\ }\textbf {\bibinfo {volume} {5}},\ \bibinfo {pages} {023024}
  (\bibinfo {year} {2023}{\natexlab{a}})}\BibitemShut {NoStop}%
\bibitem [{\citenamefont {Anisetti}\ \emph {et~al.}(2024)\citenamefont
  {Anisetti}, \citenamefont {Kandala}, \citenamefont {Scellier},\ and\
  \citenamefont {Schwarz}}]{anisetti2024frequency}%
  \BibitemOpen
  \bibfield  {author} {\bibinfo {author} {\bibfnamefont {V.~R.}\ \bibnamefont
  {Anisetti}}, \bibinfo {author} {\bibfnamefont {A.}~\bibnamefont {Kandala}},
  \bibinfo {author} {\bibfnamefont {B.}~\bibnamefont {Scellier}}, \ and\
  \bibinfo {author} {\bibfnamefont {J.~M.}\ \bibnamefont {Schwarz}},\
  }\href@noop {} {\bibfield  {journal} {\bibinfo  {journal} {Neural
  Computation}\ }\textbf {\bibinfo {volume} {36}},\ \bibinfo {pages} {596}
  (\bibinfo {year} {2024})}\BibitemShut {NoStop}%
\bibitem [{\citenamefont {Falk}\ \emph {et~al.}(2025)\citenamefont {Falk},
  \citenamefont {Strupp}, \citenamefont {Scellier},\ and\ \citenamefont
  {Murugan}}]{falk2025temporal}%
  \BibitemOpen
  \bibfield  {author} {\bibinfo {author} {\bibfnamefont {M.~J.}\ \bibnamefont
  {Falk}}, \bibinfo {author} {\bibfnamefont {A.~T.}\ \bibnamefont {Strupp}},
  \bibinfo {author} {\bibfnamefont {B.}~\bibnamefont {Scellier}}, \ and\
  \bibinfo {author} {\bibfnamefont {A.}~\bibnamefont {Murugan}},\ }\href@noop
  {} {\bibfield  {journal} {\bibinfo  {journal} {Nature Communications}\
  }\textbf {\bibinfo {volume} {16}},\ \bibinfo {pages} {2163} (\bibinfo {year}
  {2025})}\BibitemShut {NoStop}%
\bibitem [{\citenamefont {Stern}\ and\ \citenamefont
  {Murugan}(2023)}]{stern2023learning}%
  \BibitemOpen
  \bibfield  {author} {\bibinfo {author} {\bibfnamefont {M.}~\bibnamefont
  {Stern}}\ and\ \bibinfo {author} {\bibfnamefont {A.}~\bibnamefont
  {Murugan}},\ }\href@noop {} {\bibfield  {journal} {\bibinfo  {journal}
  {Annual Review of Condensed Matter Physics}\ }\textbf {\bibinfo {volume}
  {14}},\ \bibinfo {pages} {417} (\bibinfo {year} {2023})}\BibitemShut
  {NoStop}%
\bibitem [{\citenamefont {Momeni}\ \emph {et~al.}(2025)\citenamefont {Momeni},
  \citenamefont {Rahmani}, \citenamefont {Scellier}, \citenamefont {Wright},
  \citenamefont {McMahon}, \citenamefont {Wanjura}, \citenamefont {Li},
  \citenamefont {Skalli}, \citenamefont {Berloff}, \citenamefont {Onodera}
  \emph {et~al.}}]{momeni2025training}%
  \BibitemOpen
  \bibfield  {author} {\bibinfo {author} {\bibfnamefont {A.}~\bibnamefont
  {Momeni}}, \bibinfo {author} {\bibfnamefont {B.}~\bibnamefont {Rahmani}},
  \bibinfo {author} {\bibfnamefont {B.}~\bibnamefont {Scellier}}, \bibinfo
  {author} {\bibfnamefont {L.~G.}\ \bibnamefont {Wright}}, \bibinfo {author}
  {\bibfnamefont {P.~L.}\ \bibnamefont {McMahon}}, \bibinfo {author}
  {\bibfnamefont {C.~C.}\ \bibnamefont {Wanjura}}, \bibinfo {author}
  {\bibfnamefont {Y.}~\bibnamefont {Li}}, \bibinfo {author} {\bibfnamefont
  {A.}~\bibnamefont {Skalli}}, \bibinfo {author} {\bibfnamefont {N.~G.}\
  \bibnamefont {Berloff}}, \bibinfo {author} {\bibfnamefont {T.}~\bibnamefont
  {Onodera}},  \emph {et~al.},\ }\href@noop {} {\bibfield  {journal} {\bibinfo
  {journal} {Nature}\ }\textbf {\bibinfo {volume} {645}},\ \bibinfo {pages}
  {53} (\bibinfo {year} {2025})}\BibitemShut {NoStop}%
\bibitem [{\citenamefont {Dillavou}\ \emph {et~al.}(2022)\citenamefont
  {Dillavou}, \citenamefont {Stern}, \citenamefont {Liu},\ and\ \citenamefont
  {Durian}}]{dillavou2022demonstration}%
  \BibitemOpen
  \bibfield  {author} {\bibinfo {author} {\bibfnamefont {S.}~\bibnamefont
  {Dillavou}}, \bibinfo {author} {\bibfnamefont {M.}~\bibnamefont {Stern}},
  \bibinfo {author} {\bibfnamefont {A.~J.}\ \bibnamefont {Liu}}, \ and\
  \bibinfo {author} {\bibfnamefont {D.~J.}\ \bibnamefont {Durian}},\
  }\href@noop {} {\bibfield  {journal} {\bibinfo  {journal} {Physical Review
  Applied}\ }\textbf {\bibinfo {volume} {18}},\ \bibinfo {pages} {014040}
  (\bibinfo {year} {2022})}\BibitemShut {NoStop}%
\bibitem [{\citenamefont {Dillavou}\ \emph {et~al.}(2024)\citenamefont
  {Dillavou}, \citenamefont {Beyer}, \citenamefont {Stern}, \citenamefont
  {Liu}, \citenamefont {Miskin},\ and\ \citenamefont
  {Durian}}]{Dillavou2024PNAS}%
  \BibitemOpen
  \bibfield  {author} {\bibinfo {author} {\bibfnamefont {S.}~\bibnamefont
  {Dillavou}}, \bibinfo {author} {\bibfnamefont {B.~D.}\ \bibnamefont {Beyer}},
  \bibinfo {author} {\bibfnamefont {M.}~\bibnamefont {Stern}}, \bibinfo
  {author} {\bibfnamefont {A.~J.}\ \bibnamefont {Liu}}, \bibinfo {author}
  {\bibfnamefont {M.~Z.}\ \bibnamefont {Miskin}}, \ and\ \bibinfo {author}
  {\bibfnamefont {D.~J.}\ \bibnamefont {Durian}},\ }\href@noop {} {\bibfield
  {journal} {\bibinfo  {journal} {Proceedings of the National Academy of
  Sciences}\ } (\bibinfo {year} {2024})}\BibitemShut {NoStop}%
\bibitem [{\citenamefont {Altman}\ \emph {et~al.}(2024)\citenamefont {Altman},
  \citenamefont {Stern}, \citenamefont {Liu},\ and\ \citenamefont
  {Durian}}]{Altman2024}%
  \BibitemOpen
  \bibfield  {author} {\bibinfo {author} {\bibfnamefont {L.~E.}\ \bibnamefont
  {Altman}}, \bibinfo {author} {\bibfnamefont {M.}~\bibnamefont {Stern}},
  \bibinfo {author} {\bibfnamefont {A.~J.}\ \bibnamefont {Liu}}, \ and\
  \bibinfo {author} {\bibfnamefont {D.~J.}\ \bibnamefont {Durian}},\ }\href
  {\doibase 10.1103/PhysRevApplied.22.024053} {\bibfield  {journal} {\bibinfo
  {journal} {Physical Review Applied}\ }\textbf {\bibinfo {volume} {22}},\
  \bibinfo {pages} {024053} (\bibinfo {year} {2024})}\BibitemShut {NoStop}%
\bibitem [{\citenamefont {Lee}\ \emph {et~al.}(2022)\citenamefont {Lee},
  \citenamefont {Mulder},\ and\ \citenamefont {Hopkins}}]{lee2022mechanical}%
  \BibitemOpen
  \bibfield  {author} {\bibinfo {author} {\bibfnamefont {R.~H.}\ \bibnamefont
  {Lee}}, \bibinfo {author} {\bibfnamefont {E.~A.}\ \bibnamefont {Mulder}}, \
  and\ \bibinfo {author} {\bibfnamefont {J.~B.}\ \bibnamefont {Hopkins}},\
  }\href@noop {} {\bibfield  {journal} {\bibinfo  {journal} {Science Robotics}\
  }\textbf {\bibinfo {volume} {7}},\ \bibinfo {pages} {eabq7278} (\bibinfo
  {year} {2022})}\BibitemShut {NoStop}%
\bibitem [{\citenamefont {Li}\ and\ \citenamefont
  {Mao}(2024)}]{li2024training}%
  \BibitemOpen
  \bibfield  {author} {\bibinfo {author} {\bibfnamefont {S.}~\bibnamefont
  {Li}}\ and\ \bibinfo {author} {\bibfnamefont {X.}~\bibnamefont {Mao}},\
  }\href@noop {} {\bibfield  {journal} {\bibinfo  {journal} {Nature
  Communications}\ }\textbf {\bibinfo {volume} {15}},\ \bibinfo {pages} {10528}
  (\bibinfo {year} {2024})}\BibitemShut {NoStop}%
\bibitem [{\citenamefont {Falk}\ and\ \citenamefont
  {Langer}(1998)}]{FalkLanger1998}%
  \BibitemOpen
  \bibfield  {author} {\bibinfo {author} {\bibfnamefont {M.~L.}\ \bibnamefont
  {Falk}}\ and\ \bibinfo {author} {\bibfnamefont {J.~S.}\ \bibnamefont
  {Langer}},\ }\href {\doibase 10.1103/PhysRevE.57.7192} {\bibfield  {journal}
  {\bibinfo  {journal} {Physical Review E}\ }\textbf {\bibinfo {volume} {57}},\
  \bibinfo {pages} {7192} (\bibinfo {year} {1998})}\BibitemShut {NoStop}%
\bibitem [{\citenamefont {Maloney}\ and\ \citenamefont
  {Lemaître}(2006)}]{MaloneyLemaitre2006}%
  \BibitemOpen
  \bibfield  {author} {\bibinfo {author} {\bibfnamefont {C.~E.}\ \bibnamefont
  {Maloney}}\ and\ \bibinfo {author} {\bibfnamefont {A.}~\bibnamefont
  {Lemaître}},\ }\href {\doibase 10.1103/PhysRevE.74.016118} {\bibfield
  {journal} {\bibinfo  {journal} {Physical Review E}\ }\textbf {\bibinfo
  {volume} {74}},\ \bibinfo {pages} {016118} (\bibinfo {year}
  {2006})}\BibitemShut {NoStop}%
\bibitem [{\citenamefont {Kabla}\ and\ \citenamefont
  {Debr{\'e}geas}(2003)}]{KablaDebregeas2003}%
  \BibitemOpen
  \bibfield  {author} {\bibinfo {author} {\bibfnamefont {A.~J.}\ \bibnamefont
  {Kabla}}\ and\ \bibinfo {author} {\bibfnamefont {G.}~\bibnamefont
  {Debr{\'e}geas}},\ }\href {\doibase 10.1103/PhysRevLett.90.258303} {\bibfield
   {journal} {\bibinfo  {journal} {Physical Review Letters}\ }\textbf {\bibinfo
  {volume} {90}},\ \bibinfo {pages} {258303} (\bibinfo {year}
  {2003})}\BibitemShut {NoStop}%
\bibitem [{\citenamefont {Utter}\ and\ \citenamefont
  {Behringer}(2004)}]{UtterBehringer2004}%
  \BibitemOpen
  \bibfield  {author} {\bibinfo {author} {\bibfnamefont {B.}~\bibnamefont
  {Utter}}\ and\ \bibinfo {author} {\bibfnamefont {R.~P.}\ \bibnamefont
  {Behringer}},\ }\href {\doibase 10.1103/PhysRevE.69.031308} {\bibfield
  {journal} {\bibinfo  {journal} {Physical Review E}\ }\textbf {\bibinfo
  {volume} {69}},\ \bibinfo {pages} {031308} (\bibinfo {year}
  {2004})}\BibitemShut {NoStop}%
\bibitem [{\citenamefont {Keim}\ \emph {et~al.}(2019)\citenamefont {Keim},
  \citenamefont {Paulsen}, \citenamefont {Zeravcic}, \citenamefont {Sastry},\
  and\ \citenamefont {Nagel}}]{Keim2019memory}%
  \BibitemOpen
  \bibfield  {author} {\bibinfo {author} {\bibfnamefont {N.~C.}\ \bibnamefont
  {Keim}}, \bibinfo {author} {\bibfnamefont {J.~D.}\ \bibnamefont {Paulsen}},
  \bibinfo {author} {\bibfnamefont {Z.}~\bibnamefont {Zeravcic}}, \bibinfo
  {author} {\bibfnamefont {S.}~\bibnamefont {Sastry}}, \ and\ \bibinfo {author}
  {\bibfnamefont {S.~R.}\ \bibnamefont {Nagel}},\ }\href {\doibase
  10.1103/RevModPhys.91.035002} {\bibfield  {journal} {\bibinfo  {journal}
  {Rev. Mod. Phys.}\ }\textbf {\bibinfo {volume} {91}},\ \bibinfo {pages}
  {035002} (\bibinfo {year} {2019})}\BibitemShut {NoStop}%
\bibitem [{\citenamefont {Paulsen}\ and\ \citenamefont
  {Keim}(2025)}]{Paulsen25mechanical}%
  \BibitemOpen
  \bibfield  {author} {\bibinfo {author} {\bibfnamefont {J.~D.}\ \bibnamefont
  {Paulsen}}\ and\ \bibinfo {author} {\bibfnamefont {N.~C.}\ \bibnamefont
  {Keim}},\ }\href {\doibase
  https://doi.org/10.1146/annurev-conmatphys-032822-035544} {\bibfield
  {journal} {\bibinfo  {journal} {Annual Review of Condensed Matter Physics}\
  }\textbf {\bibinfo {volume} {16}},\ \bibinfo {pages} {61} (\bibinfo {year}
  {2025})}\BibitemShut {NoStop}%
\bibitem [{\citenamefont {Mungan}\ \emph {et~al.}(2019)\citenamefont {Mungan},
  \citenamefont {Sastry}, \citenamefont {Dahmen},\ and\ \citenamefont
  {Regev}}]{Mungan2019networks}%
  \BibitemOpen
  \bibfield  {author} {\bibinfo {author} {\bibfnamefont {M.}~\bibnamefont
  {Mungan}}, \bibinfo {author} {\bibfnamefont {S.}~\bibnamefont {Sastry}},
  \bibinfo {author} {\bibfnamefont {K.}~\bibnamefont {Dahmen}}, \ and\ \bibinfo
  {author} {\bibfnamefont {I.}~\bibnamefont {Regev}},\ }\href {\doibase
  10.1103/PhysRevLett.123.178002} {\bibfield  {journal} {\bibinfo  {journal}
  {Phys. Rev. Lett.}\ }\textbf {\bibinfo {volume} {123}},\ \bibinfo {pages}
  {178002} (\bibinfo {year} {2019})}\BibitemShut {NoStop}%
\bibitem [{\citenamefont {Mungan}\ and\ \citenamefont
  {Witten}(2019)}]{Mungan2019cyclic}%
  \BibitemOpen
  \bibfield  {author} {\bibinfo {author} {\bibfnamefont {M.}~\bibnamefont
  {Mungan}}\ and\ \bibinfo {author} {\bibfnamefont {T.~A.}\ \bibnamefont
  {Witten}},\ }\href {\doibase 10.1103/PhysRevE.99.052132} {\bibfield
  {journal} {\bibinfo  {journal} {Phys. Rev. E}\ }\textbf {\bibinfo {volume}
  {99}},\ \bibinfo {pages} {052132} (\bibinfo {year} {2019})}\BibitemShut
  {NoStop}%
\bibitem [{\citenamefont {Paulsen}\ and\ \citenamefont
  {Keim}(2019)}]{Paulsen2019minimal}%
  \BibitemOpen
  \bibfield  {author} {\bibinfo {author} {\bibfnamefont {J.~D.}\ \bibnamefont
  {Paulsen}}\ and\ \bibinfo {author} {\bibfnamefont {N.~C.}\ \bibnamefont
  {Keim}},\ }\href {\doibase 10.1098/rspa.2018.0874} {\bibfield  {journal}
  {\bibinfo  {journal} {Proceedings of the Royal Society A: Mathematical,
  Physical and Engineering Sciences}\ }\textbf {\bibinfo {volume} {475}},\
  \bibinfo {pages} {20180874} (\bibinfo {year} {2019})}\BibitemShut {NoStop}%
\bibitem [{\citenamefont {Keim}\ and\ \citenamefont
  {Paulsen}(2021)}]{keim2021multiperiodic}%
  \BibitemOpen
  \bibfield  {author} {\bibinfo {author} {\bibfnamefont {N.~C.}\ \bibnamefont
  {Keim}}\ and\ \bibinfo {author} {\bibfnamefont {J.~D.}\ \bibnamefont
  {Paulsen}},\ }\href@noop {} {\bibfield  {journal} {\bibinfo  {journal}
  {Science Advances}\ }\textbf {\bibinfo {volume} {7}},\ \bibinfo {pages}
  {eabg7685} (\bibinfo {year} {2021})}\BibitemShut {NoStop}%
\bibitem [{\citenamefont {Liu}\ \emph {et~al.}(2024)\citenamefont {Liu},
  \citenamefont {Teunisse}, \citenamefont {Korovin}, \citenamefont {Vermaire},
  \citenamefont {Jin}, \citenamefont {Bense},\ and\ \citenamefont {van
  Hecke}}]{Liu2024}%
  \BibitemOpen
  \bibfield  {author} {\bibinfo {author} {\bibfnamefont {J.}~\bibnamefont
  {Liu}}, \bibinfo {author} {\bibfnamefont {M.}~\bibnamefont {Teunisse}},
  \bibinfo {author} {\bibfnamefont {G.}~\bibnamefont {Korovin}}, \bibinfo
  {author} {\bibfnamefont {I.~R.}\ \bibnamefont {Vermaire}}, \bibinfo {author}
  {\bibfnamefont {L.}~\bibnamefont {Jin}}, \bibinfo {author} {\bibfnamefont
  {H.}~\bibnamefont {Bense}}, \ and\ \bibinfo {author} {\bibfnamefont
  {M.}~\bibnamefont {van Hecke}},\ }\href {\doibase 10.1073/pnas.2308414121}
  {\bibfield  {journal} {\bibinfo  {journal} {Proceedings of the National
  Academy of Sciences}\ }\textbf {\bibinfo {volume} {121}},\ \bibinfo {pages}
  {e2308414121} (\bibinfo {year} {2024})},\ \Eprint
  {http://arxiv.org/abs/https://www.pnas.org/doi/pdf/10.1073/pnas.2308414121}
  {https://www.pnas.org/doi/pdf/10.1073/pnas.2308414121} \BibitemShut {NoStop}%
\bibitem [{\citenamefont {Hopfield}(1982)}]{hopfield1982neural}%
  \BibitemOpen
  \bibfield  {author} {\bibinfo {author} {\bibfnamefont {J.~J.}\ \bibnamefont
  {Hopfield}},\ }\href@noop {} {\bibfield  {journal} {\bibinfo  {journal}
  {Proceedings of the National Academy of Sciences}\ }\textbf {\bibinfo
  {volume} {79}},\ \bibinfo {pages} {2554} (\bibinfo {year}
  {1982})}\BibitemShut {NoStop}%
\bibitem [{\citenamefont {Bitzek}\ \emph {et~al.}(2006)\citenamefont {Bitzek},
  \citenamefont {Koskinen}, \citenamefont {G\"ahler}, \citenamefont {Moseler},\
  and\ \citenamefont {Gumbsch}}]{bitzek2006structural}%
  \BibitemOpen
  \bibfield  {author} {\bibinfo {author} {\bibfnamefont {E.}~\bibnamefont
  {Bitzek}}, \bibinfo {author} {\bibfnamefont {P.}~\bibnamefont {Koskinen}},
  \bibinfo {author} {\bibfnamefont {F.}~\bibnamefont {G\"ahler}}, \bibinfo
  {author} {\bibfnamefont {M.}~\bibnamefont {Moseler}}, \ and\ \bibinfo
  {author} {\bibfnamefont {P.}~\bibnamefont {Gumbsch}},\ }\href {\doibase
  10.1103/PhysRevLett.97.170201} {\bibfield  {journal} {\bibinfo  {journal}
  {Phys. Rev. Lett.}\ }\textbf {\bibinfo {volume} {97}},\ \bibinfo {pages}
  {170201} (\bibinfo {year} {2006})}\BibitemShut {NoStop}%
\bibitem [{\citenamefont {Bengio}\ \emph
  {et~al.}(2013{\natexlab{a}})\citenamefont {Bengio}, \citenamefont
  {Courville},\ and\ \citenamefont {Vincent}}]{Bengio2013}%
  \BibitemOpen
  \bibfield  {author} {\bibinfo {author} {\bibfnamefont {Y.}~\bibnamefont
  {Bengio}}, \bibinfo {author} {\bibfnamefont {A.}~\bibnamefont {Courville}}, \
  and\ \bibinfo {author} {\bibfnamefont {P.}~\bibnamefont {Vincent}},\
  }\href@noop {} {\bibfield  {journal} {\bibinfo  {journal} {IEEE Transactions
  on Pattern Analysis and Machine Intelligence}\ }\textbf {\bibinfo {volume}
  {35}},\ \bibinfo {pages} {1798} (\bibinfo {year}
  {2013}{\natexlab{a}})}\BibitemShut {NoStop}%
\bibitem [{\citenamefont {Stern}\ \emph {et~al.}(2025)\citenamefont {Stern},
  \citenamefont {Guzman}, \citenamefont {Martins}, \citenamefont {Liu},\ and\
  \citenamefont {Balasubramanian}}]{stern2025physical}%
  \BibitemOpen
  \bibfield  {author} {\bibinfo {author} {\bibfnamefont {M.}~\bibnamefont
  {Stern}}, \bibinfo {author} {\bibfnamefont {M.}~\bibnamefont {Guzman}},
  \bibinfo {author} {\bibfnamefont {F.}~\bibnamefont {Martins}}, \bibinfo
  {author} {\bibfnamefont {A.~J.}\ \bibnamefont {Liu}}, \ and\ \bibinfo
  {author} {\bibfnamefont {V.}~\bibnamefont {Balasubramanian}},\ }\href@noop {}
  {\bibfield  {journal} {\bibinfo  {journal} {Physical Review Letters}\
  }\textbf {\bibinfo {volume} {134}},\ \bibinfo {pages} {147402} (\bibinfo
  {year} {2025})}\BibitemShut {NoStop}%
\bibitem [{\citenamefont {Goodfellow}\ \emph {et~al.}(2016)\citenamefont
  {Goodfellow}, \citenamefont {Bengio},\ and\ \citenamefont
  {Courville}}]{Goodfellow2016}%
  \BibitemOpen
  \bibfield  {author} {\bibinfo {author} {\bibfnamefont {I.}~\bibnamefont
  {Goodfellow}}, \bibinfo {author} {\bibfnamefont {Y.}~\bibnamefont {Bengio}},
  \ and\ \bibinfo {author} {\bibfnamefont {A.}~\bibnamefont {Courville}},\
  }\href@noop {} {\emph {\bibinfo {title} {Deep Learning}}}\ (\bibinfo
  {publisher} {MIT Press},\ \bibinfo {year} {2016})\BibitemShut {NoStop}%
\bibitem [{\citenamefont {LeCun}\ \emph {et~al.}(2015)\citenamefont {LeCun},
  \citenamefont {Bengio},\ and\ \citenamefont {Hinton}}]{LeCun2015}%
  \BibitemOpen
  \bibfield  {author} {\bibinfo {author} {\bibfnamefont {Y.}~\bibnamefont
  {LeCun}}, \bibinfo {author} {\bibfnamefont {Y.}~\bibnamefont {Bengio}}, \
  and\ \bibinfo {author} {\bibfnamefont {G.}~\bibnamefont {Hinton}},\
  }\href@noop {} {\bibfield  {journal} {\bibinfo  {journal} {Nature}\ }\textbf
  {\bibinfo {volume} {521}},\ \bibinfo {pages} {436} (\bibinfo {year}
  {2015})}\BibitemShut {NoStop}%
\bibitem [{\citenamefont {Mont{\'u}far}\ \emph {et~al.}(2014)\citenamefont
  {Mont{\'u}far}, \citenamefont {Pascanu}, \citenamefont {Cho},\ and\
  \citenamefont {Bengio}}]{Montufar2014}%
  \BibitemOpen
  \bibfield  {author} {\bibinfo {author} {\bibfnamefont {G.~F.}\ \bibnamefont
  {Mont{\'u}far}}, \bibinfo {author} {\bibfnamefont {R.}~\bibnamefont
  {Pascanu}}, \bibinfo {author} {\bibfnamefont {K.}~\bibnamefont {Cho}}, \ and\
  \bibinfo {author} {\bibfnamefont {Y.}~\bibnamefont {Bengio}},\ }\href@noop {}
  {\bibfield  {journal} {\bibinfo  {journal} {Advances in Neural Information
  Processing Systems}\ } (\bibinfo {year} {2014})}\BibitemShut {NoStop}%
\bibitem [{\citenamefont {Breiman}\ \emph {et~al.}(1984)\citenamefont
  {Breiman}, \citenamefont {Friedman}, \citenamefont {Olshen},\ and\
  \citenamefont {Stone}}]{Breiman1984}%
  \BibitemOpen
  \bibfield  {author} {\bibinfo {author} {\bibfnamefont {L.}~\bibnamefont
  {Breiman}}, \bibinfo {author} {\bibfnamefont {J.}~\bibnamefont {Friedman}},
  \bibinfo {author} {\bibfnamefont {R.}~\bibnamefont {Olshen}}, \ and\ \bibinfo
  {author} {\bibfnamefont {C.}~\bibnamefont {Stone}},\ }\href@noop {} {\emph
  {\bibinfo {title} {Classification and Regression Trees}}}\ (\bibinfo
  {publisher} {Wadsworth},\ \bibinfo {year} {1984})\BibitemShut {NoStop}%
\bibitem [{\citenamefont {Bengio}\ \emph
  {et~al.}(2013{\natexlab{b}})\citenamefont {Bengio}, \citenamefont
  {L{\'e}onard},\ and\ \citenamefont {Courville}}]{Bengio2013ST}%
  \BibitemOpen
  \bibfield  {author} {\bibinfo {author} {\bibfnamefont {Y.}~\bibnamefont
  {Bengio}}, \bibinfo {author} {\bibfnamefont {N.}~\bibnamefont {L{\'e}onard}},
  \ and\ \bibinfo {author} {\bibfnamefont {A.}~\bibnamefont {Courville}},\
  }\href@noop {} {\bibfield  {journal} {\bibinfo  {journal} {arXiv:1308.3432}\
  } (\bibinfo {year} {2013}{\natexlab{b}})}\BibitemShut {NoStop}%
\bibitem [{\citenamefont {Neftci}\ \emph {et~al.}(2019)\citenamefont {Neftci},
  \citenamefont {Mostafa},\ and\ \citenamefont {Zenke}}]{Neftci2019}%
  \BibitemOpen
  \bibfield  {author} {\bibinfo {author} {\bibfnamefont {E.~O.}\ \bibnamefont
  {Neftci}}, \bibinfo {author} {\bibfnamefont {H.}~\bibnamefont {Mostafa}}, \
  and\ \bibinfo {author} {\bibfnamefont {F.}~\bibnamefont {Zenke}},\
  }\href@noop {} {\bibfield  {journal} {\bibinfo  {journal} {IEEE Signal
  Processing Magazine}\ }\textbf {\bibinfo {volume} {36}},\ \bibinfo {pages}
  {61} (\bibinfo {year} {2019})}\BibitemShut {NoStop}%
\bibitem [{\citenamefont {Bengio}\ \emph {et~al.}(1994)\citenamefont {Bengio},
  \citenamefont {Simard},\ and\ \citenamefont {Frasconi}}]{Bengio1994}%
  \BibitemOpen
  \bibfield  {author} {\bibinfo {author} {\bibfnamefont {Y.}~\bibnamefont
  {Bengio}}, \bibinfo {author} {\bibfnamefont {P.}~\bibnamefont {Simard}}, \
  and\ \bibinfo {author} {\bibfnamefont {P.}~\bibnamefont {Frasconi}},\
  }\href@noop {} {\bibfield  {journal} {\bibinfo  {journal} {IEEE Transactions
  on Neural Networks}\ }\textbf {\bibinfo {volume} {5}},\ \bibinfo {pages}
  {157} (\bibinfo {year} {1994})}\BibitemShut {NoStop}%
\bibitem [{\citenamefont {Jang}\ \emph {et~al.}(2017)\citenamefont {Jang},
  \citenamefont {Gu},\ and\ \citenamefont {Poole}}]{Jang2017}%
  \BibitemOpen
  \bibfield  {author} {\bibinfo {author} {\bibfnamefont {E.}~\bibnamefont
  {Jang}}, \bibinfo {author} {\bibfnamefont {S.}~\bibnamefont {Gu}}, \ and\
  \bibinfo {author} {\bibfnamefont {B.}~\bibnamefont {Poole}},\ }\href@noop {}
  {\bibfield  {journal} {\bibinfo  {journal} {International Conference on
  Learning Representations (ICLR)}\ } (\bibinfo {year} {2017})}\BibitemShut
  {NoStop}%
\bibitem [{\citenamefont {Maddison}\ \emph {et~al.}(2017)\citenamefont
  {Maddison}, \citenamefont {Mnih},\ and\ \citenamefont {Teh}}]{Maddison2017}%
  \BibitemOpen
  \bibfield  {author} {\bibinfo {author} {\bibfnamefont {C.~J.}\ \bibnamefont
  {Maddison}}, \bibinfo {author} {\bibfnamefont {A.}~\bibnamefont {Mnih}}, \
  and\ \bibinfo {author} {\bibfnamefont {Y.~W.}\ \bibnamefont {Teh}},\
  }\href@noop {} {\bibfield  {journal} {\bibinfo  {journal} {International
  Conference on Learning Representations (ICLR)}\ } (\bibinfo {year}
  {2017})}\BibitemShut {NoStop}%
\bibitem [{\citenamefont {Hinton}\ \emph {et~al.}(2015)\citenamefont {Hinton},
  \citenamefont {Vinyals},\ and\ \citenamefont {Dean}}]{Hinton2015}%
  \BibitemOpen
  \bibfield  {author} {\bibinfo {author} {\bibfnamefont {G.}~\bibnamefont
  {Hinton}}, \bibinfo {author} {\bibfnamefont {O.}~\bibnamefont {Vinyals}}, \
  and\ \bibinfo {author} {\bibfnamefont {J.}~\bibnamefont {Dean}},\ }\href@noop
  {} {\bibfield  {journal} {\bibinfo  {journal} {arXiv:1503.02531}\ } (\bibinfo
  {year} {2015})}\BibitemShut {NoStop}%
\bibitem [{\citenamefont {Raghu}\ \emph {et~al.}(2017)\citenamefont {Raghu},
  \citenamefont {Poole}, \citenamefont {Kleinberg}, \citenamefont {Ganguli},\
  and\ \citenamefont {Sohl-Dickstein}}]{Raghu2017}%
  \BibitemOpen
  \bibfield  {author} {\bibinfo {author} {\bibfnamefont {M.}~\bibnamefont
  {Raghu}}, \bibinfo {author} {\bibfnamefont {B.}~\bibnamefont {Poole}},
  \bibinfo {author} {\bibfnamefont {J.}~\bibnamefont {Kleinberg}}, \bibinfo
  {author} {\bibfnamefont {S.}~\bibnamefont {Ganguli}}, \ and\ \bibinfo
  {author} {\bibfnamefont {J.}~\bibnamefont {Sohl-Dickstein}},\ }\href@noop {}
  {\bibfield  {journal} {\bibinfo  {journal} {International Conference on
  Machine Learning (ICML)}\ } (\bibinfo {year} {2017})}\BibitemShut {NoStop}%
\bibitem [{\citenamefont {Zu}\ and\ \citenamefont
  {Goodrich}(2025)}]{zu2025learning}%
  \BibitemOpen
  \bibfield  {author} {\bibinfo {author} {\bibfnamefont {M.}~\bibnamefont
  {Zu}}\ and\ \bibinfo {author} {\bibfnamefont {C.~P.}\ \bibnamefont
  {Goodrich}},\ }\href@noop {} {\bibfield  {journal} {\bibinfo  {journal}
  {arXiv preprint arXiv:2509.01296}\ } (\bibinfo {year} {2025})}\BibitemShut
  {NoStop}%
\bibitem [{\citenamefont {Anisetti}\ \emph
  {et~al.}(2023{\natexlab{b}})\citenamefont {Anisetti}, \citenamefont
  {Kandala},\ and\ \citenamefont {Schwarz}}]{anisetti2023emergent}%
  \BibitemOpen
  \bibfield  {author} {\bibinfo {author} {\bibfnamefont {V.~R.}\ \bibnamefont
  {Anisetti}}, \bibinfo {author} {\bibfnamefont {A.}~\bibnamefont {Kandala}}, \
  and\ \bibinfo {author} {\bibfnamefont {J.~M.}\ \bibnamefont {Schwarz}},\
  }\href {https://doi.org/10.48550/arXiv.2309.04382} {\bibfield  {journal}
  {\bibinfo  {journal} {arXiv preprint arXiv:2309.04382}\ } (\bibinfo {year}
  {2023}{\natexlab{b}})}\BibitemShut {NoStop}%
\bibitem [{\citenamefont {Banerjee}\ \emph {et~al.}(2026)\citenamefont
  {Banerjee}, \citenamefont {Falk}, \citenamefont {Gardel}, \citenamefont
  {Walczak}, \citenamefont {Mora},\ and\ \citenamefont
  {Vaikuntanathan}}]{banerjee2026learning}%
  \BibitemOpen
  \bibfield  {author} {\bibinfo {author} {\bibfnamefont {D.~S.}\ \bibnamefont
  {Banerjee}}, \bibinfo {author} {\bibfnamefont {M.~J.}\ \bibnamefont {Falk}},
  \bibinfo {author} {\bibfnamefont {M.~L.}\ \bibnamefont {Gardel}}, \bibinfo
  {author} {\bibfnamefont {A.~M.}\ \bibnamefont {Walczak}}, \bibinfo {author}
  {\bibfnamefont {T.}~\bibnamefont {Mora}}, \ and\ \bibinfo {author}
  {\bibfnamefont {S.}~\bibnamefont {Vaikuntanathan}},\ }\href@noop {}
  {\bibfield  {journal} {\bibinfo  {journal} {PRX Life}\ }\textbf {\bibinfo
  {volume} {4}},\ \bibinfo {pages} {013001} (\bibinfo {year}
  {2026})}\BibitemShut {NoStop}%
\bibitem [{\citenamefont {Henkes}\ \emph {et~al.}(2016)\citenamefont {Henkes},
  \citenamefont {Quint}, \citenamefont {Fily},\ and\ \citenamefont
  {Schwarz}}]{henkes2016rigid}%
  \BibitemOpen
  \bibfield  {author} {\bibinfo {author} {\bibfnamefont {S.}~\bibnamefont
  {Henkes}}, \bibinfo {author} {\bibfnamefont {D.~A.}\ \bibnamefont {Quint}},
  \bibinfo {author} {\bibfnamefont {Y.}~\bibnamefont {Fily}}, \ and\ \bibinfo
  {author} {\bibfnamefont {J.~M.}\ \bibnamefont {Schwarz}},\ }\href@noop {}
  {\bibfield  {journal} {\bibinfo  {journal} {Physical Review Letters}\
  }\textbf {\bibinfo {volume} {116}},\ \bibinfo {pages} {028301} (\bibinfo
  {year} {2016})}\BibitemShut {NoStop}%
\bibitem [{\citenamefont {Liu}\ \emph {et~al.}(2021)\citenamefont {Liu},
  \citenamefont {Kollmer}, \citenamefont {Daniels}, \citenamefont {Schwarz},\
  and\ \citenamefont {Henkes}}]{liu2021spongelike}%
  \BibitemOpen
  \bibfield  {author} {\bibinfo {author} {\bibfnamefont {K.}~\bibnamefont
  {Liu}}, \bibinfo {author} {\bibfnamefont {J.~E.}\ \bibnamefont {Kollmer}},
  \bibinfo {author} {\bibfnamefont {K.~E.}\ \bibnamefont {Daniels}}, \bibinfo
  {author} {\bibfnamefont {J.~M.}\ \bibnamefont {Schwarz}}, \ and\ \bibinfo
  {author} {\bibfnamefont {S.}~\bibnamefont {Henkes}},\ }\href@noop {}
  {\bibfield  {journal} {\bibinfo  {journal} {Physical Review Letters}\
  }\textbf {\bibinfo {volume} {126}},\ \bibinfo {pages} {088002} (\bibinfo
  {year} {2021})}\BibitemShut {NoStop}%
\bibitem [{\citenamefont {Henkes}\ \emph {et~al.}(2011)\citenamefont {Henkes},
  \citenamefont {Fily},\ and\ \citenamefont {Marchetti}}]{henkes2011active}%
  \BibitemOpen
  \bibfield  {author} {\bibinfo {author} {\bibfnamefont {S.}~\bibnamefont
  {Henkes}}, \bibinfo {author} {\bibfnamefont {Y.}~\bibnamefont {Fily}}, \ and\
  \bibinfo {author} {\bibfnamefont {M.~C.}\ \bibnamefont {Marchetti}},\
  }\href@noop {} {\bibfield  {journal} {\bibinfo  {journal} {Physical Review
  E—Statistical, Nonlinear, and Soft Matter Physics}\ }\textbf {\bibinfo
  {volume} {84}},\ \bibinfo {pages} {040301} (\bibinfo {year}
  {2011})}\BibitemShut {NoStop}%
\end{thebibliography}%

%========== Supplemental Material 开始（粘在 \end{document} 之前） ==========%

\clearpage
\onecolumngrid   % 从 twocolumn 切换为单栏显示 SI

% 重置计数器并改成 S 编号
\setcounter{secnumdepth}{3}
\setcounter{equation}{0}
\setcounter{figure}{0}
\setcounter{table}{0}

\makeatletter
\renewcommand{\thefigure}{S\@arabic\c@figure}
\renewcommand{\thetable}{S\@Roman\c@table}
\renewcommand{\theequation}{S\@arabic\c@equation}
\makeatother

\section{Supplemental Material}

\begin{figure}[h]
	\centering
	\includegraphics[width=0.9\textwidth]{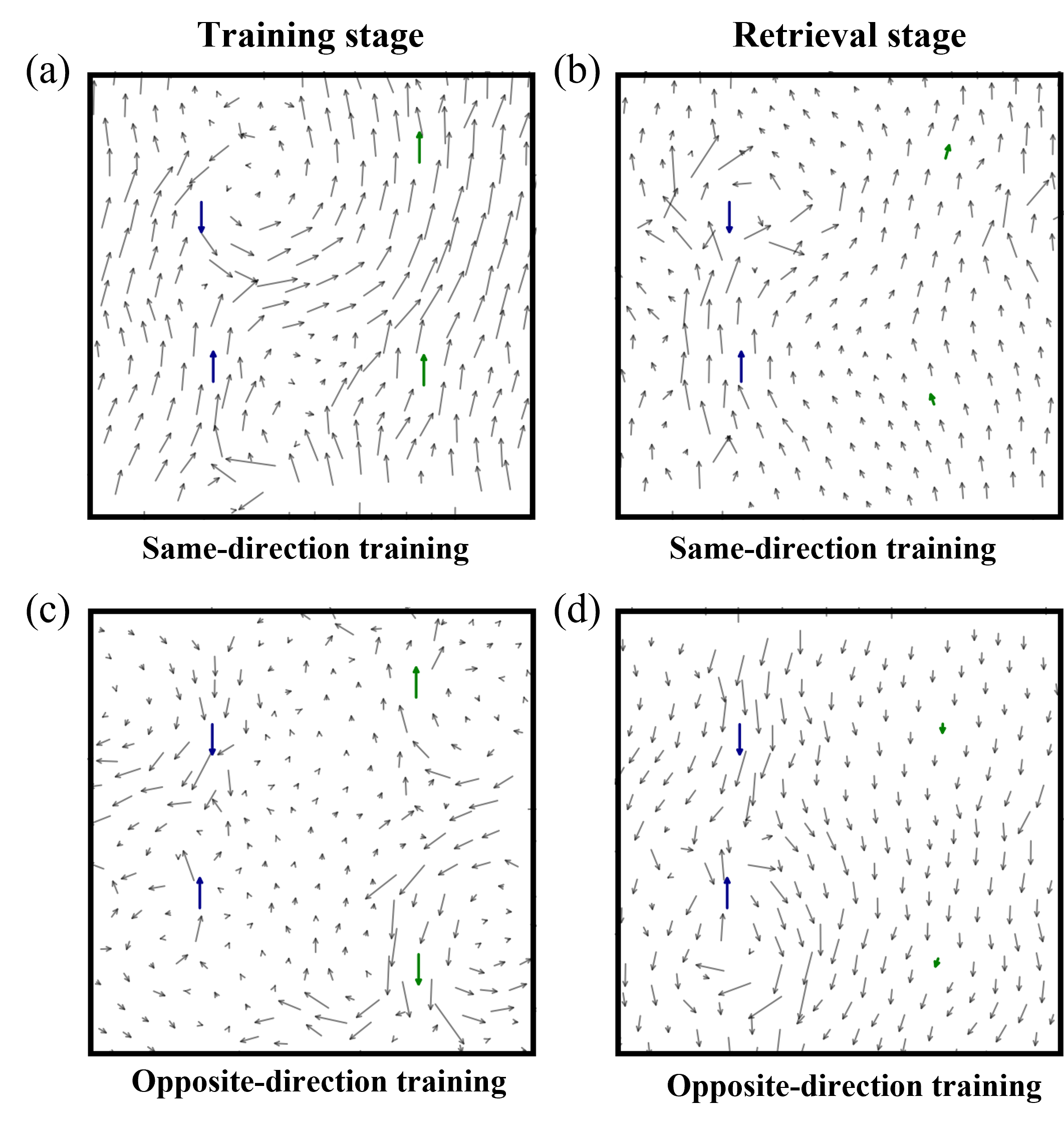}
	\caption{{\it Easy task: displacement fields at maximum input displacement (training vs.\ retrieval).}
	Gray arrows show instantaneous particle displacements at the half-cycle point where the driven inputs reach their maximum displacement. Blue arrows indicate the imposed input-particle displacements. 
	In the training-stage panels (a,c), green arrows indicate the \emph{imposed} output-particle displacements (outputs are actively driven). 
	In the retrieval-stage panels (b,d), outputs are undriven and the green arrows indicate the output particles' \emph{measured} instantaneous displacement directions. 
	(a) Training stage, same-direction output-drive protocol: snapshot from the final training cycle after rearrangements have stabilized. 
	(b) Retrieval stage for the same trained system (inputs driven; outputs undriven). 
	(c) Training stage, opposite-direction output-drive protocol: snapshot from the final stabilized training cycle. 
	(d) Retrieval stage for the opposite-direction-trained system (inputs driven; outputs undriven).}
	\label{fig:easyDisplacementField_S1}
\end{figure}

\begin{figure*}[h]
	\centering
	\includegraphics[width=0.9\textwidth]{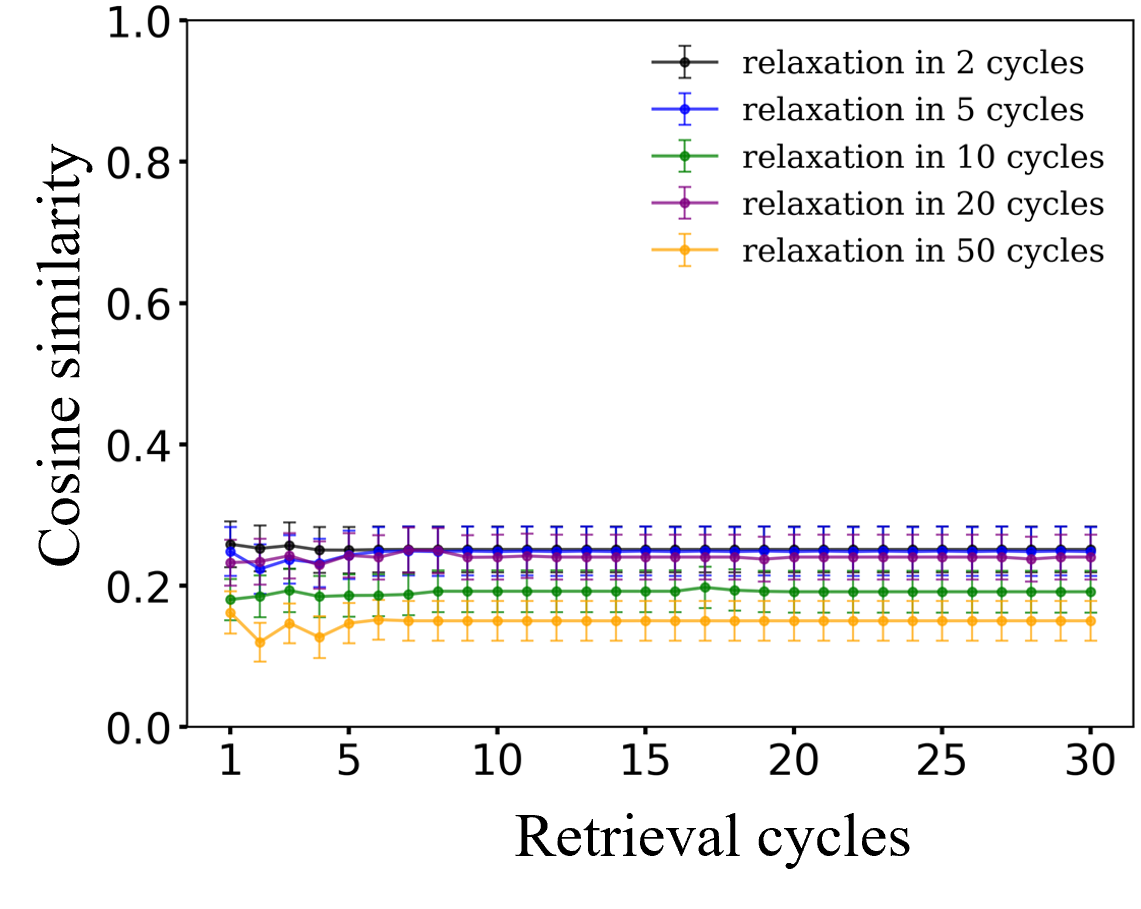}
	\caption{{\it Hard task: effect of intermittent relaxation interval.}
Cosine similarity $S_{\mathrm{dir}}$ versus retrieval cycle number for the hard task using intermittent relaxation during training,
where every $N_r$ retrieval cycles the outputs are left undriven in the $N_r$th cycle. Curves show $N_r=2,5,10,20,50$.}
	\label{fig:hardtask1relaxation}
\end{figure*}

\begin{figure*}[t]
	\centering
	\includegraphics[width=0.95\textwidth]{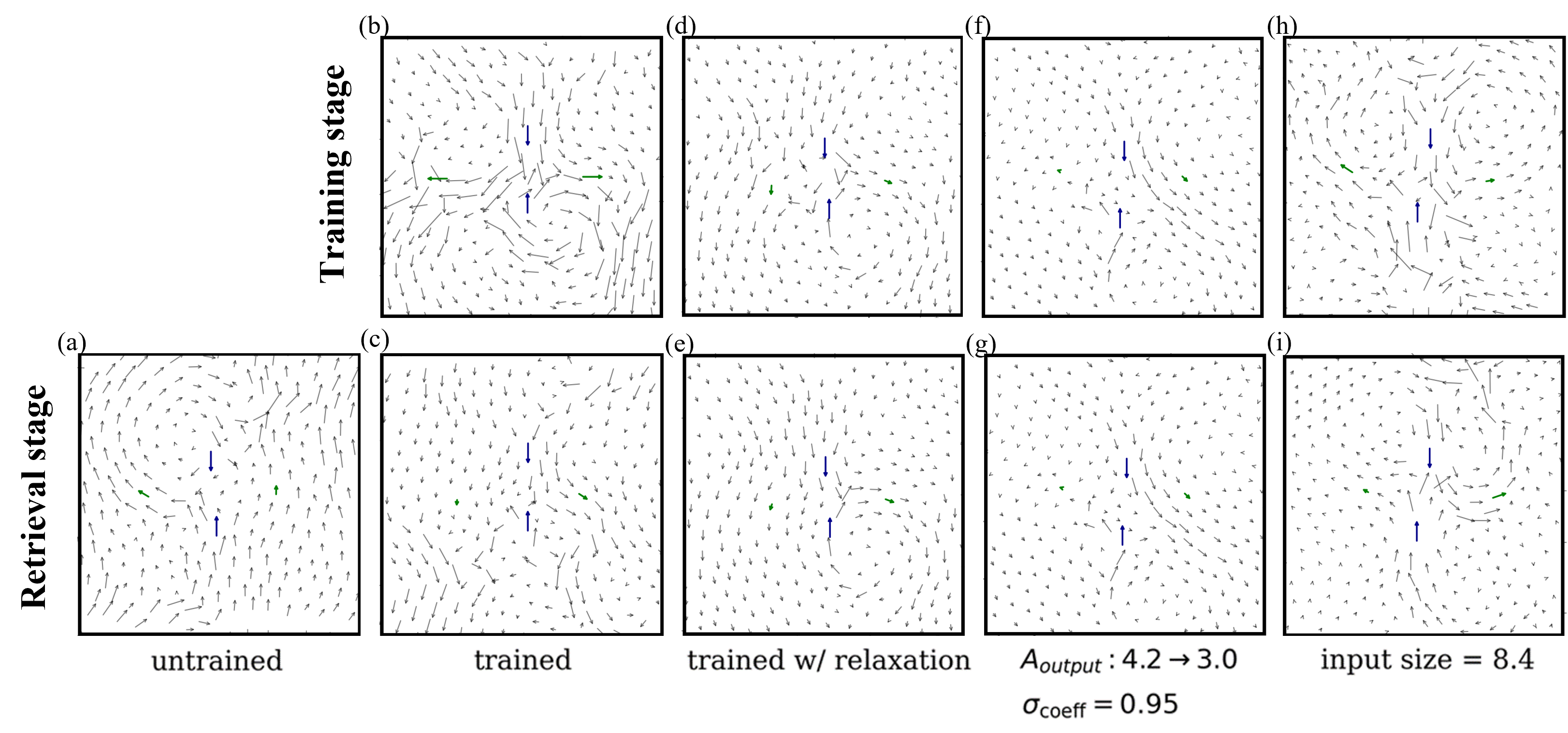}
	\caption{{\it Quadrupolar task: displacement fields at maximum input displacement (training vs.\ retrieval) under different protocols.}
	Gray arrows show instantaneous particle displacements at the half-cycle point where the driven inputs reach their maximum displacement; blue arrows indicate the imposed input-particle displacements. All snapshots are taken from the \textit{final} cycle of the corresponding phase (training or retrieval), after rearrangements have stabilized. The top row shows the training stage; the bottom row shows the retrieval stage. Columns correspond to: untrained; baseline trained; trained with intermittent relaxation; trained with relaxation plus annealed output drive ($A_{\mathrm{out}}:4.2\!\rightarrow\!3.0$) and $\sigma_{\mathrm{coeff}}=0.95$; and the same protocol with enlarged input particles (input size $=8.4$). Green arrows mark the output particles. In the baseline-trained training-stage panel, green arrows indicate the \textit{imposed} output-particle driving directions (outputs actively driven). In the other training-stage panels, the final training cycle is a relaxation cycle (outputs undriven), so green arrows indicate the outputs' \textit{measured} instantaneous displacement directions. In all retrieval-stage panels, outputs are undriven and green arrows likewise indicate the outputs' \textit{measured} instantaneous displacement directions.}
	\label{fig:quadrupoleDisplacementField_S3}
\end{figure*}

\begin{figure*}[h]
	\centering
	\includegraphics[width=0.9\textwidth]{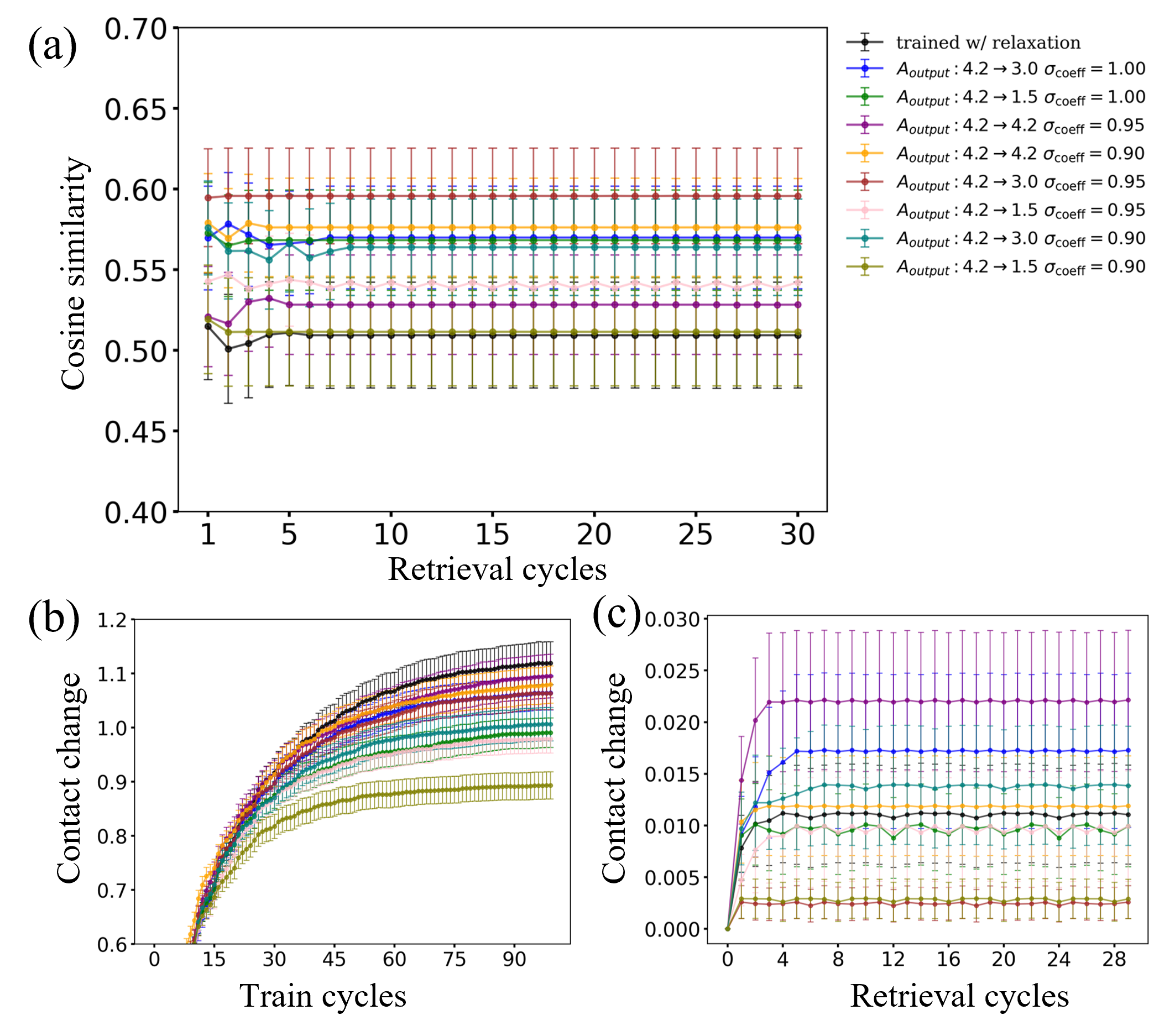}
	\caption{{\it Quadrupolar task: parameter sweep.}
(a) Directional cosine similarity $S_{\mathrm{dir}}$ versus retrieval cycle number for the quadrupolar geometry (cf.\ Fig.~\ref{fig:type3}). All curves use intermittent relaxation during training (every 5 training cycles the outputs are left undriven). The legend lists combined parameter variations: during training the output driving amplitude is linearly ramped from $A_{\mathrm{out}}=4.2$ in the first training cycle to the final value indicated after the arrow ($3.0$ or $1.5$; $4.2\!\rightarrow\!4.2$ denotes constant amplitude), and the Lennard--Jones length prefactor is set to $\sigma_{\mathrm{coeff}}=1.00$, $0.95$, or $0.90$ as labeled.
(b) Total contact change $\mathrm{TCC}(t)$ versus training cycle number for the same parameter sets as in (a).
(c) $\mathrm{TCC}(t)$ versus retrieval cycle number for the same parameter sets as in (a).
Error bars indicate variability across 100 independent realizations.}
	\label{fig:quadrupole_parameters}
\end{figure*}

% \begin{figure}[!htbp]
%     \centering
%     \includegraphics[width=0.9\textwidth]{SI/drive3.png}
%     \caption{{\it drive 3 input particles} (a) Schematic of
%             training (b) Schematic of testing (c) cosine similarity} 
%     \label{fig:drive3}
% \end{figure}

%========== Supplemental Material 结束 ==========%

\end{document}